\documentclass[11pt,a4paper]{article}
\pdfoutput=1

\usepackage{amsfonts}
\usepackage{amssymb}
\usepackage{mathrsfs}
\usepackage{amsmath,amssymb}
\usepackage{subeqn}
\usepackage{bbm}
\usepackage[bookmarks=true]{hyperref}
\usepackage[nosort]{cite}
\usepackage[bulletsep]{collect}

\ifx\hypersetup\sadfkjashdfkxja\else
\hypersetup{pdftitle={Dual Superconformal Symmetry from AdS5 x S5 Superstring Integrability}}
\hypersetup{pdfauthor={Niklas Beisert, Riccardo Ricci, Arkady A. Tseytlin, Martin Wolf}}
\hypersetup{pdfsubject={Mathematical Physics}}
\hypersetup{pdfkeywords={Integrability, T Duality, Lax Connection}}
\hypersetup{plainpages=false}
\hypersetup{bookmarksnumbered=true}
\hypersetup{pdfstartview=FitH}
\hypersetup{pdfpagemode=None}
\hypersetup{colorlinks=false}
\hypersetup{citebordercolor={.5 1 .5}}
\hypersetup{urlbordercolor={.5 1 1}}
\hypersetup{linkbordercolor={1 .7 .7}}
\fi

\makeatletter\renewcommand{\section}{\@startsection
{section}{1}{\z@}{-2.5ex plus -1ex minus
    -.2ex}{2.3ex plus .2ex}{\centering\large\bf\mathversion{bold}}}

\makeatletter\renewcommand{\subsection}{\@startsection{subsection}{2}{\z@}{-3.25ex
plus -1ex minus
   -.2ex}{1.5ex plus .2ex}{\centering\bf\mathversion{bold}}}

\makeatletter\renewcommand{\subsubsection}{\@startsection{subsubsection}{3}{-2.45ex}{-3.25ex
plus -1ex minus -.2ex}{1.5ex plus .2ex}{\centering\bf\mathversion{bold}}}

\makeatletter\renewcommand{\paragraph}{\@startsection{paragraph}{4}{\z@}%
                                    {0.8ex \@plus1ex \@minus.2ex}%
                                    {-.5em}%
                                    {\normalfont\normalsize\bfseries\mathversion{bold}}}

\renewcommand{\thesection}{\arabic{section}.}

\numberwithin{paragraph}{section}
\renewcommand\theparagraph {\S\thesection\@arabic\c@paragraph.\kern-8pt}

\numberwithin{equation}{section}

\renewcommand*\l@section{\@dottedtocline{1}{0em}{1.5em}}
\renewcommand*\l@subsubsection{\@dottedtocline{4}{3.8em}{3.2em}}

\renewcommand\tableofcontents{%
    \section*{\large\contentsname
        \@mkboth{%
          \MakeUppercase\contentsname}{\MakeUppercase\contentsname}}%
       {\baselineskip=15pt plus 2pt minus 1pt
    \@starttoc{toc}}%
}

\DeclareFontFamily{U}{rsf}{}
\DeclareFontShape{U}{rsf}{m}{n}{
  <5> <6> rsfs5 <7> <8> <9> rsfs7 <10-> rsfs10}{}
\DeclareMathAlphabet\Scr{U}{rsf}{m}{n}

\renewcommand{\d}{\mathrm{d}}
\newcommand{\im}{\mathrm{i}}
\newcommand{\eu}{\mathrm{e}}

\newcommand{\IR}{\mathbbm{R}}
\newcommand{\IZ}{\mathbbm{Z}}

\newcommand{\cC}{\mathscr{C}}

\newcommand{\cP}{\mathscr{P}}

\newcommand{\CN}{\mathcal{N}}
\newcommand{\CM}{\mathcal{M}}
\newcommand{\CO}{\mathcal{O}}
\newcommand{\CU}{\mathcal{U}}

\newcommand{\fg}{\mathfrak{g}}
\newcommand{\fh}{\mathfrak{h}}

\def\la{\label}
\def\foot{\footnote}

\newcommand{\rf}[1]{(\ref{#1})}
\def \s {\sigma}

\def \ooo {{(0)}} 
\def \two {{(2)}}
\def \td {\tilde}
\def \vp {\varphi}
\def \hj {{\rm j}}
\def \hJ {{\rm J}}
\def \aA {{\cal V}}
\def \BB {{\rm B}}

\begin{document}

\begin{titlepage}

\setcounter{page}{0}
\renewcommand{\thefootnote}{\fnsymbol{footnote}}

\begin{flushright}
AEI--2008--044\\
Imperial--TP--RR--01/2008\\[.5cm]
\end{flushright}

\begin{center}

{\LARGE\textbf{\mathversion{bold}
Dual Superconformal Symmetry\\
from  $\mathrm{AdS}_5\times S^5$  Superstring
Integrability}\par}

\vspace*{.8cm}

{\large
Niklas Beisert$^a$,
Riccardo Ricci$^{b,}$\footnote{Also at the Institute for Mathematical Sciences, Imperial College, London, U.K.}, Arkady A. Tseytlin$^{b,}$\footnote{Also at the Lebedev Institute, Moscow, Russia.}
and Martin Wolf$^{b}$ \footnote{{\it E-mail addresses:\/}
{\ttfamily nbeisert@aei.mpg.de,~$\{$r.ricci, a.tseytlin, m.wolf$\}$@imperial.ac.uk}}
}

\vspace*{.8cm}

{\it $^{a}$ Max-Planck-Institut f\"ur Gravitationsphysik\\
Albert-Einstein-Institut\\
Am M\"uhlenberg 1, 14476 Potsdam, Germany}\\[.5cm]
{\it $^{b}$ Theoretical Physics Group\\
The Blackett Laboratory, Imperial College London\\
Prince Consort Road, London SW7 2AZ, United Kingdom}\\[.5cm]

\vspace*{.3cm}

{\bf Abstract}

\end{center}

\vspace*{-.3cm}

\begin{quote}

We discuss $2d$ duality transformations in the classical AdS$_5\times S^5$
superstring and their effect on the integrable structure.
T-duality along four directions in the Poincar\'e parametrization
of AdS$_5$ maps the bosonic part of the superstring action into itself.
On the bosonic level, this duality may be understood as a symmetry
of the first-order (phase space) system of equations for the coset components
of the current. The associated Lax connection is invariant modulo the action of 
an $\mathfrak{so}(2,4)$-automorphism. We then show that this symmetry
extends to the full superstring, provided one supplements the
transformation of the bosonic components of the current  with a transformation
on the fermionic ones. At the level of the action, this symmetry can be
seen by combining the bosonic duality transformation with a similar one applied
to part of the fermionic superstring coordinates. As a result, the full superstring
action is mapped  into itself, albeit in a different $\kappa$-symmetry gauge.
One implication is that the  dual model has the same  superconformal
symmetry group as the original one, and this may be seen as a consequence of the
integrability of the superstring. The invariance  of the Lax connection
under the duality implies a map on the full set of conserved charges that should
interchange some of the Noether (local) charges with hidden (non-local) ones and
vice versa.

\vfill
\noindent July 21, 2008

\end{quote}

\setcounter{footnote}{0}\renewcommand{\thefootnote}{\arabic{thefootnote}}

\end{titlepage}

\tableofcontents


\section{Introduction}

The integrability of string theory in AdS$_5\times S^5$  holds major promise
of a complete solution for its spectrum and thus for the spectrum
of dimensions of gauge invariant operators of the dual
$\CN=4$ supersymmetric Yang--Mills (SYM) theory. The bosonic 
AdS$_5\times S^5$ sigma model is classically integrable
being based on the  coset $SO(2,4)/SO(1,4)\times SO(6)/SO(5)$
\cite{Luscher:1977rq,Mandal:2002fs}. That property extends  to the full Green--Schwarz (GS) 
superstring model based on the supercoset $PSU(2,2|4)/(SO(1,4)\times SO(5))$ \cite{Metsaev:1998it},
as follows from its special $\kappa$-symmetry structure \cite{Bena:2003wd}.

The integrability formally implies the existence of an infinite number of conserved charges
and thus of an infinite-dimensional hidden symmetry algebra of the $2d$ string sigma model.
In addition to the obvious kinematic superconformal  $PSU(2,2|4)$ symmetry ``seen'' by a
point-particle limit of the string and corresponding to the standard local Noether conserved
charges, there are also hidden ``stringy'' symmetries and the associated charges.

The integrable structure of a $G/H$  coset sigma model is naturally
defined on a phase space, i.e.\ in terms of a family of currents or a  Lax connection
whose flatness implies the complete system of first-order differential equations
for the components  of the current. This system includes the Maurer--Cartan part as
well as a ``dynamical'' part which follows from the standard coset sigma model Lagrangian.

Like in many similar examples (Maxwell equations, etc.), the first-order system in a sense
is more general than the usual coset model: while it does not directly follow from a local
Lagrangian, it may lead to different ``dual'' local Lagrangian representations. The latter 
are found by a ``phase space reduction'' procedure -- by solving part
of the first-order equations and substituting the solution into the rest
which may then follow from a ``dual'' Lagrangian. Various sigma model dualities
can be understood in that way (see, e.g., \cite{Zakharov:1973pp,Nappi:1979ig,Fridling:1983ha,Fradkin:1984ai}). The dual sigma models will
then be classically equivalent (i.e.\ their classical solutions will be directly  related)
and will share the same integrable structure, i.e.\ hidden symmetries.

Those of such classical duality transformations which can be implemented by a change of
variables in a path integral formulation of the theory  can then be extended to the quantum
level. They then define quantum-dual sigma models \cite{Fridling:1983ha,Fradkin:1984ai} in the sense that there 
is a prescription relating certain quantum correlators in one theory to certain correlators
in the dual theory. The well-known  example is the standard $2d$ scalar  duality or T-duality
in which case the path integral transformation relating the two dual theories may be formulated
in terms of gauging an Abelian isometry \cite{Buscher:1987qj,Rocek:1991ps}.

In general, such sigma model dualities do not respect manifest global symmetries of the theory,
that is, symmetries ``seen'' by a point-like string. For example, the usual T-duality in $x$
direction relates the models  with target-space metrics $\d s^2 =\d r^2 + a^2(r)\d x^2$ 
and $\d s^2 = \d r^2 + a^{-2} (r) \d x^2$, so that in the case of $S^2$ when $a=\sin(r)$ and
$x$ is compact, the first model  has $SO(3)$ symmetry while the second only $SO(2)$. The two
sigma models are still classically equivalent, i.e.\ they share the same first-order formulation
and integrable structure. Put differently, the corresponding $2d$ field theories (i.e.\ string
models as opposed to their point-particle truncations) are, in a sense, equally 
symmetric.\foot{This is  similar to the fate of space-time supersymmetry under T-duality: it
remains a symmetry of the underlying conformal field theory but may become non-locally realized
(see, e.g., \cite{Alvarez:1995zr,Sfetsos:1995ac,Curtright:1996ig}).}

\

A special case  is  provided by the  AdS$_n$ sigma model in Poincar\'e coordinates 
($a=\eu^{r}$ in the above example corresponds to AdS$_2$), where the formal T-duality along all
$(n-1)$ translational directions combined with a simple coordinate transformation $r\mapsto -r$
gives back  an  equivalent sigma model on the ``dual'' AdS$_n$ space \cite{Kallosh:1998ji}. This means that 
the original and the dual  models happen to have equivalent global Noether symmetries 
$SO(2,n-1)$ but realized  on different (dual) sets of variables. This ``T-self-duality'' 
property of AdS$_5$  was used in \cite{Alday:2007hr,Alday:2007he} to construct classical solutions for open strings
related to the strong-coupling limit of gluon scattering amplitudes (see also \cite{Kruczenski:2007cy}). 
It appears also to be related to the ``dual conformal symmetry'' of maximally
helicity violating (MHV) amplitudes observed at weak coupling \cite{Drummond:2006rz,Drummond:2007aua,Drummond:2007cf,Drummond:2007au,Drummond:2008aq,Bern:2006ew,Bern:2007ct,Bern:2008ap}.\footnote{The
T-duality  on string side seems to be intimately connected with the relation between  
the gluon scattering amplitudes and Wilson loops at strong \cite{Alday:2007hr,Alday:2007he,Komargodski:2008wa} 
(see also \cite{Polyakov:1998ju,Polyakov:2000fk}) and weak \cite{Drummond:2006rz,Drummond:2007aua,Drummond:2007cf,Drummond:2007au,Drummond:2008aq,Bern:2006ew,Bern:2007ct,Bern:2008ap,Brandhuber:2007yx,McGreevy:2008zy} coupling.}

Since the two dual models  are classically equivalent and also share the same integrable
structure, the local Noether charges of the dual model should be related to hidden (non-local)
charges of the original model and vice versa. One can indeed express the Lax connection in
terms of either original or dual variables and thus, in principle, relate the charges in the
two pictures \cite{Ricci:2007eq}.\footnote{For previous work on the relation between local and non-local
charges under T-duality in a different context see, e.g., \cite{Hatsuda:2006ts}.} Given that
the sigma model has an infinite number of hidden charges, the true significance of this
``doubling'' of a particular subset of them, i.e.\ of the Noether symmetry charges, still remains
to be understood.

As was suggested in \cite{Ricci:2007eq}, and this will be one of the aims of the present paper, one should
be able to extend this construction of the flat currents and associated charges to the dual of
the full AdS$_5\times S^5$ superstring model. In the process, we will show explicitly that the
Lax connections of the original and the dual sigma models are not independent but equivalent 
(in particular, one does not get ``doubling'' of conserved charges).

\

Very recently, it was suggested \cite{SokTalk,KorTalk,Drummond:2008vq} that the dual conformal symmetry of perturbative
$\CN=4$ SYM theory MHV amplitudes \cite{Drummond:2006rz,Drummond:2007aua,Drummond:2007cf,Drummond:2007au,Drummond:2008aq} has an extension to the full ``dual superconformal
symmetry'' if one considers the full set of supergluon amplitudes. Simultaneously, it was
suggested \cite{BerTalk} that, if the T-duality transformation of the AdS$_5\times S^5$ superstring
action along the four translational directions of AdS$_5$ \cite{Kallosh:1998ji} is followed by a similar 
$2d$ duality transformation applied to part of the fermionic worldsheet variables
(corresponding to the Poincar\'e supersymmetry generators $Q$ but not $\bar Q$),
then the dual action will take the original form, i.e.\ it will be again equivalent to the
AdS$_5\times S^5$ supercoset GS action. Thus, the dual Noether charges will again generate 
the full superconformal group $PSU(2,2|4)$.

Below we shall extend the discussion in \cite{Ricci:2007eq} by showing that in the bosonic AdS$_{n}$ sigma
model the particular $2d$ duality corresponding to the T-duality can be realized as a discrete 
symmetry of the phase space equations,\foot{The usual $2d$ scalar duality $\d x\mapsto{*\d}x$
or $j\mapsto{* j}$ can be viewed as a phase space transformation (see, e.g., \cite{Giveon:1994fu,Alvarez:1994dn}) that 
exchanges momenta $\partial_\tau x $ with some combinations of coordinates $\partial_\s x $ 
(or $x_{k-1}-x_k$ in discrete mode representation).} i.e.\ of the first-order system of equations
for the components of the $SO(2,n-1)/SO(1,n-1)$ coset current. Furthermore, this duality can be
reinterpreted as a (spectral parameter dependent) automorphism of the global 
$\mathfrak{so}(2,n-1)$ symmetry algebra \eqref{eq:algauto}. 
This fact then makes the original and the dual
integrable structures equivalent in a precise manner.

We shall also consider the full AdS$_5\times S^5$ superstring case in the same $\kappa$-symmetry
gauge as in \cite{Kallosh:1998ji}\foot{This ``S-gauge'' \cite{Kallosh:1998nx,Pesando:1998fv,Metsaev:2000yf,Metsaev:2000yu} is a natural choice for a comparison with
the boundary gauge theory as it corresponds to setting the fermionic components associated to the
superconformal generators $S,\bar S$ to zero. It naturally complements the parametrization of the
bosonic part of the supercoset in terms of the coordinates corresponding to the translational $P$, 
dilational $D$ and the $R$-symmetry $SU(4)$-generators.} in which the dual action is quadratic in 
the fermions and explicitly perform the bosonic duality transformation in the Lax connection.
We shall then follow the idea of \cite{BerTalk} and supplement the bosonic duality by a fermionic duality
transformation to discover that the resulting action (which is again quadratic in the fermions)
can be identified with the original AdS$_5\times S^5$ superstring action written in a different 
(complex) $\kappa$-symmetry gauge (used previously in \cite{Roiban:2000yy}).

We shall show that the necessity of the fermionic duality transformation becomes transparent in
the first-order formulation in terms of the supercoset current: the discrete symmetry of the phase
space system which corresponded to the bosonic T-duality should be extended to act also on the
fermionic components of the currents in order to make it possible to identify the original and
the dual Lax connections upon an  action of an  automorphism of the $\mathfrak{psu}(2,2|4)$
superalgebra \eqref{eq:superauto2}.\foot{In general, the duality as a symmetry of the first-order system should be
understood modulo a choice of $G/H$ coset representative, i.e.\ local $H$-symmetry gauge, and a
choice of $\kappa$-symmetry gauge (and also modulo certain analytic continuation).} This not only 
implies that the resulting ``T-dual'' model should have the full ``dual'' superconformal symmetry
but should also allow, in principle, to establish the duality isomorphism on the full infinite set
of conserved charges. In that sense, the dual superconformal symmetry may be viewed as a 
consequence of integrability.\foot{Conformal invariance and integrability are expected to promote
this classical symmetry to a quantum one. To attempt to relate it to dual conformal symmetry of
the boundary gauge theory one should combine its action on the ``bulk'' string coordinates with
action on the vertex operators (inserted at the boundary or an IR brane \cite{Alday:2007hr,Alday:2007he}) so that the gluon
scattering amplitudes as computed on the string side become invariant.}

From a broader perspective, this ``T-duality'' is just one particular symmetry of the first-order
system for the AdS$_5\times S^5$ superstring. One may consider also other duality transformations
that will lead to equivalent classical systems; for example, one  may mix the fermionic and bosonic
dualities in a different order, etc. These more general transformations are also worth studying
(though not all of them may have path integral, i.e.\ quantum counterparts) as they may further
clarify the structure  of this integrable theory. The special feature of ``T-duality'' is that it
preserves the maximal possible global symmetry group. The existence of such transformation appears
to be deeply rooted in the structure of the superconformal algebra: the possibility to choose 
the translational subalgebra as maximal Abelian subalgebra in $SO(2,4)$ ($[P_a,P_b]=0$) together
with its $\CN=4$ Poincar\'e supersymmetry counterpart ($\{Q^{i\alpha}, Q^{j\beta}\}=0=
[P_a,Q^{i\alpha}]$); the invariant meaning of this $2d$ duality transformation is that it acts on
the associated four bosonic and eight fermionic $2d$ fields.

Let us emphasize once more the point (already mentioned in \cite{Ricci:2007eq}) that  given conserved
Noether charges in the original sigma model, we may express them in terms of the dual variables 
and thus, get a collection of (possibly non-local) conserved charges in the dual model. The 
existence of an additional  set of conserved Noether charges in the dual model which are local
in the dual variables and thus non-local in the original variables means that they must originate
from some hidden conserved charges in the original model, and this may be viewed as a consequence
of its integrability. This explains the title of the present paper.

\

The structure of the paper is as follows:

In Sec.~\ref{sec:BCMD}, we make some general comments on $2d$ duality transformations in the group
$G$ and the coset $G/H$ bosonic sigma models.

In Sec.~\ref{sec:review}, we review the structure of the AdS$_5\times S^5$ superstring sigma model
using the supercoset construction. We shall present the equations of motion in first-order
form and describe two important families of flat currents implying integrability of this model.
Then we shall specialize the discussion to the standard choice of the basis of generators 
of the superconformal algebra, choose a parametrization of the supercoset adapted to Poincar\'e
coordinates in AdS$_5$ and fix a particular $\kappa$-symmetry gauge in which the sigma model
action takes a simple form.

In Sec.~\ref{sec:BTduality}, we follow \cite{Kallosh:1998ji} and transform the superstring  action by applying $2d$
duality (T-duality) to four scalar fields corresponding to translational directions of AdS$_5$.
We shall show how to express the flat Lax connection in terms of the dual variables which 
implies integrability of the dual model. We shall then consider the first-order system of
equations for the current and, first ignoring the fermions, show that the  T-duality
transformation can be understood as a symmetry of this system and of the corresponding Lax
connection. This means that one does not get two copies of the integrable structure  but rather
an automorphism on the space of conserved charges \eqref{eq:algauto}.

In Sec.~\ref{sec:last}, we show, following the suggestion of \cite{BerTalk}, that combining the bosonic
duality with a similar duality transformation on half of fermions present in the
$\kappa$-symmetry gauge fixed action  one gets back the same AdS$_5\times S^5$ superstring
action but written in a different $\kappa$-symmetry gauge (and modulo a certain analytic
continuation). This implies recovering after the duality the full global superconformal
symmetry. The combined action of bosonic and fermionic duality transformations will then
be understood more abstractly as a symmetry of the first-order system and the Lax connection
which manifests itself as an automorphism of the superconformal algebra: under its action,
original and dual Lax connections get identified \eqref{eq:superauto2} (see also Table 5.1).

The first three appendices contain our notation and some technical details, while the last one
contains some comments on the construction of conserved charges in the case of closed string
worldsheet. 

\

\section{Bosonic coset models and their dualities}\label{sec:BCMD}

Before turning to the AdS$_5\times S^5$ superstring and specifics of T-duality, let us make few
general comments  about classical sigma model dualities in the context of the bosonic $G/H$ coset model.

Let us start with the principal chiral model (PCM) based on
\begin{equation}
 L\ =\ \tfrac{1}{2}\mbox{tr} ( j\wedge {*j} ),\qquad\mbox{with}
 \qquad  j\ =\ g^{-1} \d g\qquad\mbox{and}\qquad g\ \in\ G.
\end{equation}
The corresponding equations of motion written in first-order form are
\begin{equation}
 \d j + j\wedge j\ =\ 0\qquad\mbox{and}\qquad\d{*j}\ =\ 0. \la{ps}
\end{equation}
These  two  equations \rf{ps} follow from the condition of flatness of the following
family of currents or Lax connection, with $z$ as a complex spectral parameter:
\begin{equation} \la{lax}
 \hj(z)\ =\ a j + b{*j},\qquad\mbox{with}\qquad
 a\ =\ -\tfrac{1}{4}(z-z^{-1})^2\qquad\mbox{and}\qquad b\ =\ \tfrac{1}{4}(z^2-z^{-2}).
\end{equation}
The standard second-order PCM equation is found by solving the first (Maurer--Cartan)
equation in \rf{ps} as $j=g^{-1}\d g$ and then substituting the solution into the second equation.

Instead, we may construct a dual model \cite{Zakharov:1973pp} by first solving the second equation in \rf{ps} as
$j={*\d}\chi$, where $\chi\in\fg:={\rm Lie}(G)$ is the dual field and substituting this into the 
Maurer--Cartan equation. The resulting equation $\d{*\d}\chi-\d\chi\wedge\d\chi =0$ then follows
from the dual Lagrangian
\begin{equation} \la{gg}
 \td L\ =\ \tfrac{1}{2}\mbox{tr}(\d\chi \wedge {*\d}\chi  + 
  \tfrac{2}{3} \chi\,\d \chi \wedge \d \chi).
\end{equation}
Note that if we write $g=\eu^{\eta}$ then for small $\eta$ the relation between $\eta$ and $\chi$
is the same as the usual $2d$ scalar duality, $\d\eta={*\d}\chi$. If we introduce the dual current
$\td j=\d\chi$, then the first-order system  for the dual model will be\foot{We assume Minkowski 
signature on the worldsheet.}
$\d{*\td j}-\td j\wedge\td j=0$ and $\d\td j=0$, i.e.\ it will be equivalent to the original one
\rf{ps} under $j\mapsto{* \td j}$. This transformation will leave the Lax connection \rf{lax}
invariant provided we supplement it by $z\mapsto\eu^{\frac\pi4\im}z$ and an overall rescaling by
`$\im$'. Note that the Noether symmetries of the original and dual sigma models here are different.

The model \rf{gg} is sometimes called ``pseudo-dual'' \cite{Nappi:1979ig,Curtright:1994be,Sarisaman:2007dm} to reflect the fact that it is
not quantum-equivalent to the original PCM. To construct the quantum-equivalent ``non-Abelian
dual'' of the PCM \cite{Fridling:1983ha,Fradkin:1984ai}, one has to start with
\begin{equation} 
 \bar L\ =\ \mbox{tr}\bigl[\tfrac{1}{2}j\wedge{*j}+\vp (\d j +j\wedge j)\bigr], \la{kk}   
\end{equation}
where $\vp\in\fg$  plays the role of a Lagrange multiplier, and subsequently integrate out $j$.
The resulting dual model will again be classically equivalent to the PCM.\foot{Indeed, while 
the classical equations that follow from \rf{kk} will involve an extra field $\vp$, one can 
easily see that they imply \rf{ps}. One finds that ${*j}=-\nabla\vp$  and $\d j+j\wedge j=0$
but then $\nabla^2=0$ which leads to $\d{*j}=0$.} It has equivalent integrable structure but
(after we solve for $j$) will have smaller Noether symmetry.

Let us now turn to the case of the $G/H$ symmetric space coset model given by
\begin{equation} \la{opp}
 L\ =\ \tfrac{1}{2}\mbox{tr}(j_\two\wedge{*j}_\two),\qquad\mbox{with}\qquad
 j\ =\ g^{-1} \d g\ =\ j_\ooo + j_\two\ \equiv\ A +j_\two,
\end{equation}
where we split the current according to the $\IZ_2$-decomposition of the Lie  algebra
$\fg\cong \fg_\ooo + \fg_\two\equiv  \fh + \fg_\two$. The corresponding first-order system may
be written as ($\nabla:=\d + A$)
\begin{equation} \la{coo}
 \d A + A\wedge A  + j_\two\wedge j_\two\ =\ 0,\qquad\nabla  j_\two\ =\ 0\qquad\mbox{and}\qquad
 \nabla{*j}_\two\ =\ 0, 
\end{equation}
where the first two equations are the $\fh$ and $\fg_\two$ components of the Maurer--Cartan
equation. These equations follow from the flatness of a Lax connection similar to the one 
in \rf{lax}
\begin{equation} \la{cax}
 \hj(z)\ =\ A + a j_\two + b {*j}_\two, \quad\mbox{with}\quad
  a\ =\ \tfrac 12(z^2+z^{-2})\quad\mbox{and}\quad b\ =\ -\tfrac 12(z^2-z^{-2}).
\end{equation}
Here we observe a formal duality symmetry of this phase space system and its integrable structure
under $j_\two\mapsto\im{*j}_\two$ and  $z\mapsto\eu^{\frac\pi4\im}z$. 
To relate the coset fields, we may define a non-local map 
$g\mapsto\td g$ such that $(g^{-1}\d g)_\two = *(\td g^{-1}\d\td g)_\two$.\foot{One may consider
similar formal Hodge duality transformation also in the superstring case \cite{Xiong:2007if}.}

One may also consider here an analog of the non-Abelian duality transformation in the PCM
that can be performed at the path integral level by adding the Maurer--Cartan equations with the 
Lagrange multiplier fields   to the action  and then solving for the current components
(for an example in the $S^2$-case, see \cite{Mikhailov:2005sy}).

In addition to this formal symmetry, there may be  other ``dualities'', i.e.\ linear
transformations of the current components that map this first-order  system into itself and
respect its integrable structure. The T-duality that we are going to discuss below in the special
case of AdS$_5\cong SO(2,4)/SO(1,4)$ is one of them that has a remarkable property of being
a ``self-duality'': it maps the system into an equivalent one with the {\it same} $SO(2,4)$ global
symmetry.

\

\section{Review of \texorpdfstring{${\rm AdS}_5\times S^5$}{AdS5 x S5} superstring sigma model}\label{sec:review}

We will begin this section with a summary of the supercoset formulation of the type
IIB superstring action on AdS$_5\times S^5$. We will then move on to the discussion
of some aspects of its classical integrability by reviewing the construction of flat
currents. We will also  explicitly construct the Noether currents for the supercoset
model in the parametrization adapted to the standard basis of the superconformal group.
These currents will later be our starting point in the construction of a family of
gauge-invariant flat currents for the T-dual model.

\subsection{Superstring action}\label{sec:action-eom}

As was shown in \cite{Metsaev:1998it}, the type IIB Green--Schwarz superstring action on
AdS$_5\times S^5$ can be understood as a sigma model-type action on the coset superspace
\begin{equation}\label{eq:supercoset}
 G/H\ =\ PSU(2,2|4)/(SO(1,4)\times SO(5))
\end{equation}
with the bosonic part being
\begin{equation}
 SO(2,4)/SO(1,4)\times SO(6)/SO(5)\ \cong\ {\rm AdS}_5\times S^5.
\end{equation}
The coset \eqref{eq:supercoset} admits a $\IZ_4$-grading in the sense that the
subgroup $H=SO(1,4)\times SO(5)$ of $G=PSU(2,2|4)$ arises
as the fixed point set of an order $4$ automorphism
of $G$ \cite{Berkovits:1999zq}. Concretely,  this means that at the Lie algebra
 level $\fg:=\mbox{Lie}(G)$ we have ($m,n=0,\ldots,3$)
\begin{equation}\label{eq:algsplit}
 \fg\ \cong\ \bigoplus_{m=0}^3\fg_{(m)},
 \quad\mbox{with}\quad\fg_{(0)}\ \equiv\ \fh\ :=\ \mbox{Lie}(H)
 \quad\mbox{and}\quad [\fg_{(m)},\fg_{(n)}\}\ \subset\ \fg_{(m+n)}.
\end{equation}
Here  $\fg_{(0)}$ and $\fg_{(2)}$ are generated by bosonic generators while
$\fg_{(1)}$ and $\fg_{(3)}$ by fermionic ones, respectively (for
more details, see Sec.~\ref{sec:sconfalg} below).

To define  the superstring action, we  consider  the map $g:\Sigma\to G$, where
$\Sigma$ is a  worldsheet surface
(with an arbitrary Lorentzian $2d$  metric) and introduce the current
\begin{equation}\label{eq:flatcurrent}
 j\ =\ g^{-1}\d g\ =\ j_{(0)} + j_{(1)}+j_{(2)}+j_{(3)},
 \quad
 \mbox{with}
 \quad j_{(0)}\ =:\ A\ \in\ \fh\quad\mbox{and}\quad j_{(m)}\ \in\ \fg_{(m)}.
\end{equation}
 The  dynamical $2d$  fields (string coordinates) will take values
in the coset superspace $G/H:=\{gH\,|\,g\in G\}$. The action that describes them should
 simultaneously be invariant under the global (left) $G$-transformations of the form
\begin{subequations}
\begin{equation}\label{eq:gtrafo}
 g\ \mapsto\ g_0g\qquad\mbox{for}\qquad g_0\ \in\ G,
\end{equation}
and the local (right) $H$-transformations of
the form
\begin{equation}\label{eq:htrafo}
 g\ \mapsto\ gh\qquad\mbox{for}\qquad h\ \in\ H.
\end{equation}
\end{subequations}
By construction, the current $j$ is invariant under \eqref{eq:gtrafo}. 
Under \eqref{eq:htrafo}, the $A$-part of $j$ in \rf{eq:flatcurrent} 
transforms  as a connection, $A\mapsto h^{-1}Ah+h^{-1}\d h$, while the $j_{(m)}$s with 
$m=1,2,3$ transform covariantly, $j_{(m)}\mapsto h^{-1} j_{(m)} h$.

The superstring action can  be written as a sum of kinetic and Wess--Zumino (WZ) terms
\cite{Metsaev:1998it,Berkovits:1999zq,Roiban:2000yy},
\begin{equation}\label{eq:action1}
 S\ =\ -\tfrac{T}{2}\int_\Sigma\mbox{str}\,\big[j_{(2)}\wedge{*j_{(2)}}+\kappa j_{(1)}\wedge
 j_{(3)}\big] ,
\end{equation}
where $T=\frac{\sqrt{\lambda}}{2\pi}$ is the string tension, `$*$' is the Hodge star on $\Sigma$
and   `str' denotes the supertrace on
$\fg$ compatible with the $\IZ_4$-grading,
\begin{equation}
 \mbox{str}(V_mV_n)\ =\ 0,\qquad\mbox{for}\qquad V_m\ \in\
 \mathfrak{g}_{(m)}
 \qquad\mbox{and}\qquad m+n\ \not=\ 0 \ {\rm mod}\  4.
\end{equation}
The $\kappa$-symmetry condition requires that $\kappa=\pm 1$; in what follows we shall
assume that (the opposite  sign choice is related by parity transformation on $\Sigma$)
\begin{equation} 
 \kappa\ =\ 1. 
\end{equation}
Note that regardless of the requirement of $\kappa$-symmetry, the superstring action 
\rf{eq:action1} is integrable \cite{Bena:2003wd} only for the same choice of $\kappa =\pm 1$.
This is not totally surprising since (i) the bosonic coset model is classically integrable
\cite{Luscher:1977rq} and (ii) it is local $\kappa$-symmetry that relates bosons to fermions and thus
extends this property to the fermionic GS generalization of the bosonic coset model.\foot{The
same applies also to similar lower-dimensional GS models constructed in \cite{Polyakov:2004br}.}

\subsection{Equations of motion}

Starting with the Maurer--Cartan equation for the current \eqref{eq:flatcurrent}
\begin{equation}\label{eq:MC}
 \d j+j\wedge j\ =\ 0
\end{equation}
and splitting  it   according to the $\IZ_4$-grading of the algebra
gives (cf.~\rf{coo})
\begin{equation}\label{iden}
 \begin{aligned}
  \d A+A\wedge A +j_{(1)}\wedge j_{(3)}+j_{(2)}\wedge j_{(2)}+j_{(3)}\wedge j_{(1)}\ &=\ 0,\\
  \nabla j_{(1)}+j_{(2)}\wedge j_{(3)}+j_{(3)}\wedge j_{(2)}\ &=\ 0,\\
  \nabla j_{(2)}+j_{(1)}\wedge j_{(1)}+j_{(3)}\wedge j_{(3)}\ &=\ 0,\\
  \nabla j_{(3)}+j_{(1)}\wedge j_{(2)}+j_{(2)}\wedge j_{(1)}\ &=\ 0.
 \end{aligned}
\end{equation}
Here, for $\alpha$ being a Lie algebra valued $p$-form on $\Sigma$, we defined
\begin{equation}
 \nabla\alpha\ :=\ \d\alpha+A\wedge\alpha-(-)^p\alpha\wedge A .
\end{equation}
The variation of \eqref{eq:action1} over $g$  together with \eqref{iden} then yields
the following field equations:
\begin{equation}\label{eom}
 \begin{aligned}
  \nabla{*j_{(2)}}+ j_{(3)}\wedge j_{(3)}-j_{(1)}\wedge j_{(1)}\ &=\ 0,\\
  j_{(2)}\wedge(j_{(1)}+ {*j_{(1)}})+(j_{(1)}+ {*j_{(1)}})\wedge j_{(2)}\ &=\ 0,\\
  j_{(2)}\wedge(j_{(3)}- {*j_{(3)}})+(j_{(3)}- {*j_{(3)}})\wedge j_{(2)}\ &=\ 0.
 \end{aligned}
\end{equation}
Eqs.~\rf{iden} and \rf{eom} constitute the full system of superstring
equations in first-order form, i.e.\ the equations for the superalgebra valued
one-form $j$. This system is invariant under the bosonic $H$-gauge  transformations
and  the fermionic $\kappa$-gauge symmetry\foot{Under $\kappa$-symmetry we have
$\delta_\kappa j = \d \epsilon + [j, \epsilon]$ where $\epsilon= \epsilon_+ + \epsilon_- $ 
is a certain combination of self-dual and anti-self-dual fermionic vector parameters with 
$ j_{(2)}$ and also $\delta (\sqrt {-g} g^{ab}) \sim  \epsilon_+ j_{(1)} + \epsilon_- j_{(3)}$
(for details see \cite{Metsaev:1998it,Alday:2005gi,Arutyunov:2004yx,Grigoriev:2007bu,Mikhailov:2007xr}).} (and also $2d$ reparame\-trizations). This invariance
will be important to keep in mind when discussing the duality transformations later on.

Eqs.~\rf{eom}, understood as second-order equations on $g$, imply and also are implied by
the conservation condition
\begin{equation} \la{noe}
 \d{*J_{{_N}}}\ =\ 0
\end{equation}
for the Noether current $J_{_N}$ associated with the global $G$-symmetry \eqref{eq:gtrafo} of the
action. As follows from the action \eqref{eq:action1}, $J_{_N}$ is given by
\begin{equation}\label{eq:noether}
 J_{_N}\ =\ g\big[j_{(2)}-\tfrac{1}{2}{*(j_{(1)}-j_{(3)})}\big]g^{-1}.
\end{equation}
Note that, like the action itself, $J_{_N}$ is invariant under the $H$-gauge transformations
\eqref{eq:htrafo}.

To study the ``physical'' string dynamics, one needs to take care of the gauge symmetries.
The local $H$-symmetry \rf{eq:htrafo} can be fixed by making a particular choice of the coset
representative (i.e.\ the explicit choice of $g$ in terms of the independent string coordinates);
one should also choose a $\kappa$-symmetry gauge. We will discuss some particular choices below.
In general, one needs also to add the equations of motion for the $2d$ metric (i.e.\ the Virasoro
constraints) and to fix a $2d$ reparametrization gauge but this will not be required for the aims
of the present paper.\foot{For a discussion of integrability of the superstring model with the 
gauges fixed and the Virasoro constraints imposed, see \cite{Alday:2005gi,Arutyunov:2004yx,Grigoriev:2007bu,Mikhailov:2007xr}.}

\subsection{One-parameter families of flat currents}\label{sec:1pFC}

As was shown in \cite{Bena:2003wd}, the $\IZ_4$-grading of the above $G/H$ supercoset allows
for the construction of one-parameter families of flat currents.\foot{See \cite{Young:2005jv}
for the extension to $\IZ_m$-graded coset (super)spaces.} These (related) families of flat currents
allow in turn for the construction of infinitely many non-local conserved charges \`a la L\"uscher
and Pohlmeyer \cite{Luscher:1977rq}.

Indeed, one may verify that the following combination of the components of the current in
\eqref{eq:flatcurrent}
\begin{equation}\label{eq:1PF}
 \hj(z)\ =\ A + z\,j_{(1)}+\tfrac{1}{2}(z^2+z^{-2})\,j_{(2)}+z^{-1}\, j_{(3)}
 -\tfrac{1}{2}(z^2-z^{-2})\,{*j_{(2)}},
\end{equation}
where $z$ is a complex spectral parameter \cite{Beisert:2005bm} so that $\hj(1)=j$, satisfies the flatness
condition
\begin{equation} \la{fla1}
 \d \hj(z)+\hj(z)\wedge \hj(z)\ =\ 0.
\end{equation}
And vice versa, imposing this flatness condition leads to the  full system  \rf{iden}, \rf{eom}
of first-order equations for the current $j$.

Note that,  like $j$ itself, the  family of currents  $\hj(z)$  is  not invariant
under the $H$-gauge transformations \eqref{eq:htrafo}, i.e.\ it depends on a particular
choice of representative of the coset $G/H$. At the same time, starting with $\hj$ one may 
also construct another family of flat currents
\begin{equation}\label{eq:ginvflatcur}
 \hJ(z)\ :=\ g \big[ \hj(z)-\hj(1)\big]  g^{-1}\ =\ g   \big[ \hj(z)-j \big]g^{-1}\ =
 \ g\hj(z)g^{-1}+g\d g^{-1}
\end{equation}
that is invariant with respect to \eqref{eq:htrafo}.\footnote{The bosonic part of the current
$\hJ(z)$ is analogous to the Lax connection of the PCM given in Eq.~\eqref{lax} and will reduce
to it in the limit $G/H\rightarrow G$ in which $H$ becomes trivial.} That requires, however,
the explicit use of $g$, related to $j$ by $j= g^{-1}\d g$, so that $\hJ$ itself is non-local
once expressed in  terms of $j$. Expanding $\hJ(z)$ in powers of $w:=-2\log(z)$ around zero 
(i.e.\ around $z=\pm1$), we get
\begin{equation}\label{eq:exp1}
 \hJ(z)\ =\ \sum_{k=0}^\infty w^k c_k\ =\ w c_1+\CO(w^2),
\end{equation}
where $c_1$ is, in fact, the Hodge dual of the Noether current \eqref{eq:noether},
\begin{equation}\label{eq:exp2}
 c_1\ =\ {*J_{_N}}\ = \ g\big[{*j}_{(2)}-\tfrac{1}{2}{(j_{(1)}-j_{(3)})}\big]g^{-1}.
\end{equation}
Hence, the flatness of $\hJ(z)$
\begin{equation}\label{eq:flat2}
 \d \hJ(z)+\hJ(z)\wedge \hJ(z)\ =\ 0
\end{equation}
implies the conservation law $\d{*J_{_N}}=0$ and thus also the second-order equations of
motion \rf{eom} for the superstring.

Since $\hJ(z)$ is flat, we may write
\begin{equation}\label{charge}
 \hJ(z)\ =\ W^{-1}(z)\d W(z)\quad \Longrightarrow\quad
 W(z;\sigma,\tau;\sigma_0,\tau_0)\ =\ P\exp\left(\int_\cC \hJ(z)\right)\!,
\end{equation}
where $\cC$ is a contour on the worldsheet  $\Sigma$ running from some reference point
$(\sigma_0,\tau_0)$ to $(\sigma,\tau)$ and `$P$' is the path-ordering symbol. Provided
that appropriate boundary conditions at spatial infinity can be chosen, one can use the
path-ordered exponential $W$ to build an infinite number of conserved non-local charges
\cite{Luscher:1977rq}\footnote{For more details, see, e.g., the review in \cite{Dolan:1983bp}.} for
the superstring.

Let us point out that the flatness conditions \rf{fla1} and \rf{eq:flat2} are invariant
under formal  $G$-gauge transformations (with parameter ${\rm g}$), so that, e.g., $\hJ(z)$
is unique up to
\begin{equation}\label{gah}
 \hJ(z)\ \mapsto\ \hJ'(z)\ =\ {\rm g}^{-1}\hJ(z){\rm g}+{\rm g}^{-1}\d {\rm g},
 \qquad\mbox{for}\qquad{\rm g}\ \in\ G.
\end{equation}
The power series expansion of $\hJ'(z)$ in $w=-2\log(z)$ around zero is then\foot{Here
${\rm g}$ is assumed not to depend on the spectral parameter; if it does, such a transformation
may be interpreted as a ``dressing'' transformation.}
\begin{equation}
 \hJ'(z)\ =\ {\rm g}^{-1}\d {\rm g}+w\, {\rm g}^{-1}{*J_{_{N}}}{\rm g}+\CO(w^2)
\end{equation}
so that, to zeroth order in $w$, Eq.~\eqref{eq:flat2} is automatically satisfied
while to first order we again find $\d{*J_{{_N}}}=0$ or the equations of motion. Below we
shall use this gauge freedom to achieve a particularly simple form of the currents suitable
for expressing them in terms of the T-dual variables.

\subsection{Standard choice of the superconformal algebra basis}\label{sec:sconfalg}

Let us now  make a specific choice of the basis  of generators of the superconformal algebra $\mathfrak{g}=\mathfrak{psu}(2,2|4)$ adapted to  the Poincar\'e parametrization of AdS$_5$
and thus a natural one for  comparison with boundary conformal gauge theory in $\IR^{1,3}$
(see Appendix \ref{sec:SCA} for more details):
\begin{equation}\label{eq:4dgen}
 \mathfrak{psu}(2,2|4)\ =\ \mbox{span}\big\{P_a,L_{ab},K_a,D,{R_i}^j\,|\,Q^{i\alpha},
 \bar Q_i^{\dot\alpha},S_i^{\alpha},\bar S^{i\dot\alpha}\big\},
\end{equation}
where $a,b=0,\ldots,3$, $\alpha,\beta=1,2$, $\dot\alpha,\dot\beta=\dot1,\dot2$ and $i,j=1,\ldots,4$.
Here $P$ represents translations, $L$ - Lorentz rotations, $K$ - special conformal transformations,
$D$ - dilatations and $R$ - the $SU(4)$-symmetry while $Q$ and $\bar Q$ are the Poincar\'e
supercharges and $S$ and $\bar S$ their superconformal partners. We shall assume that the 
generators $P$, $K$, $D$, $L$ are Hermitian while ${R_i}^j=-({R_j}^i)^\dagger$, \
$Q^{i\alpha}=(\bar Q^{\dot\alpha}_i)^\dagger$ and $S_i^\alpha=(\bar S^{i\dot\alpha})^\dagger$.
Later on, we will also make use of the standard vector/bi-spinor index identification
$\{a\}=\{\alpha\dot\beta\}$ as discussed in Appendix \ref{sec:gammamatrix}.

In terms of these generators, the $\IZ_4$-splitting \eqref{eq:algsplit} is then given as
\begin{equation}\label{eq:explicitsplitting}
\begin{aligned}
 \fh\ &=\ \mbox{span}\big\{ \tfrac{1}{2}(P_a-K_a),L_{ab},R_{(ij)}\big\},\\
 \fg_{(1)}\ &=\ \mbox{span}\big\{ \tfrac{1}{2}(Q^{i\alpha}+C^{ij}S_j^\alpha),\tfrac{1}{2}
                 (\bar Q_i^{\dot\alpha}+C_{ij}\bar S^{j\dot\alpha})\big\},\\
 \fg_{(2)}\ &=\ \mbox{span}\big\{ \tfrac{1}{2}(P_a+K_a),D,R_{[ij]}\big\},\\
 \fg_{(3)}\ &=\ \mbox{span}\big\{\tfrac{-\im}{2}(Q^{i\alpha}-C^{ij}S_j^\alpha), \tfrac{\im}{2}
                 (\bar Q_i^{\dot\alpha}-C_{ij}\bar S^{j\dot\alpha})\big\}.
\end{aligned}
\end{equation}
Here, $R_{ij}:=C_{ik} R_j^{\ k}$ and in $R_{(ij)}$ ($R_{[ij]}$) the parentheses (square
brackets) mean normalized (anti-)symmetrization. The constant matrix $C_{ij}$ is an
$Sp(4)$-metric and has the properties\footnote{Here, `$*$' denotes complex conjugation
and $\epsilon_{ijkl}$ is totally anti-symmetric with $\epsilon_{1234}=1$.}
\begin{equation}
  C_{ij}\ =\ -C_{ji}\ =:\ \tfrac{1}{2}\epsilon_{ijkl}C^{kl},\qquad 
  C_{ij}\ =\ (C^{ij})^*\qquad\mbox{and}\qquad C_{ik}C^{jk}\ =\ \delta^j_i, 
\end{equation}
and it may be interpreted  as a charge conjugation acting on $SU(4)$-spinors.
We should stress that the particular choice of $C_{ij}$ will  not matter in the end, since
physical quantities will not depend on it.

\subsection{Poincar\'e parametrization  of supercoset representative}

Writing  the current \eqref{eq:flatcurrent} in the basis \rf{eq:4dgen}, we get
\begin{equation}\label{eq:cartan}
\begin{aligned}
 j\ &=\ j_{_{P_a}}P_a+j_{_{L_{ab}}}L_{ab}+j_{_{K_a}}K_a+j_{_D}D+j_{_{{R_i}^j}}{R_i}^j\ \\
  &\kern3cm+\ \im (j_{_{Q^{i\alpha}}}Q^{i\alpha}-
    j_{_{Q_i^{\dot\alpha}}}\bar Q_i^{\dot\alpha}+ j_{_{S_i^\alpha}}S_i^\alpha
  - j_{_{S^{i\dot\alpha}}}\bar S^{i\dot\alpha}),
\end{aligned}
\end{equation}
where the factor of `$\im$' in front of the fermionic part
was chosen  to make $j$ skew-Hermitian.
Our aim now is to  find the explicit  form  of these components
in the parametrization of the supercoset corresponding to the  Poincar\'e form of the
AdS$_5\times S^5$ metric:
\begin{equation}\label{eq:ini}
 \d s^2\ =\ -\tfrac{1}{2}Y^2\d X_{\alpha\dot\beta}\d X^{\dot\beta\alpha}+
          \tfrac{1}{4Y^2}\,\d Y_{ij}\d Y^{ij}.
\end{equation}
Here, $(X,Y)=(X^{\dot\alpha\beta},Y^{ij})$ represent the 10 independent
bosonic coordinates (see also Appendix \ref{sec:gammamatrix})\footnote{Here, 
$\epsilon_{12}=-\epsilon_{21}=\epsilon_{\dot1\dot2}=-\epsilon_{\dot2\dot1}=1$.}
\begin{equation}\label{eq:abb1}
\begin{aligned}
 &\kern1.5cm X_{\alpha\dot\beta}\ :=\ \sigma^a_{\alpha\dot\beta} X_a\ =\
 \epsilon_{\alpha\gamma}\epsilon_{\dot\beta\dot\delta}X^{\dot\delta\gamma}\ =\ 
  (X_{\dot\beta\alpha})^* ,\\
 &Y_{ij}\ =\ -Y_{ji}\ =\ \tfrac{1}{2}\epsilon_{ijkl}Y^{kl}\ =\ 
  (Y^{ij})^* \qquad\mbox{and}\qquad Y^2\ := \ \tfrac{1}{4}Y_{ij} Y^{ij} .
\end{aligned}
\end{equation}
The coset representative $g\in G$ of $[g]\in G/H$ adapted to the metric
\eqref{eq:ini} may  be chosen  as\foot{This corresponds to a particular gauge fixing of the
local $H$-symmetry  (cf.~\rf{eq:explicitsplitting}).
A similar parametrization was used  in \cite{Metsaev:2000yf,Metsaev:2000yu}.}
\begin{subequations}\label{eq:rep1}
\begin{equation}
 g(X,Y,\Theta)\ =\ B(X,Y)\, \eu^{-F(\Theta)},
\end{equation}
with
\begin{equation}
\begin{aligned}
 B(X,Y)\ &=\ \eu^{\im X^{\dot\alpha\beta}P_{\beta\dot\alpha}}\,\eu^{\im\log(Y)\, D}\,\Lambda (Y)
          \ =\ \eu^{\im X^{\dot\alpha\beta}P_{\beta\dot\alpha}}\,Y^{\im D}\,\Lambda (Y),\\
 F(\Theta)\ &=\  \im\big[ C_{ij}\epsilon_{\alpha\beta}(\theta^{i\alpha}_+Q^{j\beta}+
  \theta_-^{i\alpha}C^{jk}S_k^\beta)-C^{ij}\epsilon_{\dot\alpha\dot\beta}(\bar\theta_{+i}^{\dot\alpha}
  \bar Q_j^{\dot\beta}+\bar\theta_{-i}^{\dot\alpha}C_{jk}\bar S^{k\dot\beta})\big],
\end{aligned}
\end{equation}
where\footnote{Notice that $\Lambda^{-1}=\Lambda^\dagger$ and $\det \Lambda =1$
upon using Eqs.~\eqref{eq:abb1} and  the reality condition.}
\begin{equation}\label{eq:lambda}
 \Lambda(Y) \ =\ ({\Lambda^i}_j)\ :=\  \tfrac{1}{Y}(C^{ik} Y_{kj}).
\end{equation}
\end{subequations}
Here, $\Theta=(\theta_\pm^{i\alpha},\bar\theta_{\pm i}^{\dot\alpha})$ represents
the 32 independent fermionic coordinates  satisfying the following reality condition:
\begin{equation} \la{rea}
 \theta^{i\alpha}_\pm\ =\ (\bar\theta^{\dot\alpha}_{\pm i})^\dagger. 
\end{equation}
Then the current  may be written as
\begin{subequations}
\begin{equation}\la{ja}
 j\ =\ g^{-1}\d g\ =\ \eu^{F} j_B\,\eu^{-F}+\eu^{F}\d \eu^{-F}\ =\ j_B(X,Y)+j_F(X,Y,\Theta),
\end{equation}
where
\begin{equation}
j_B\ :=\ j(X,Y,\Theta=0)\ =\ B^{-1}\d B.
\end{equation}
\end{subequations}
A simple calculation shows that
\begin{equation}\label{eq:boscur}
 j_B\ =\ \underbrace{\im\,Y\d X^{\dot\alpha\beta}}_{=:\,j_{B_{P_{\beta\dot\alpha}}}}
 P_{\beta\dot\alpha}+\underbrace{\tfrac{\im}{Y}\d Y}_{=:\,j_{B_D}} D +
  \underbrace{2\im {(\Lambda^{-1})^i}_k\d{\Lambda^k}_j}_{=:\,j_{B_{{R_i}^j}}} {R_i}^j.
\end{equation}
As reviewed in Appendix \ref{sec:CR}, the fermionic part of  the current
can be expressed as \cite{Kallosh:1998zx,Claus:1998yw}
\begin{equation}\label{eq:unfixedcurrent}
 j_F\ =\  -\tfrac{\sinh(\mathcal{M}) }{\mathcal{\CM}}\,\nabla F
          -2\left[F, \tfrac{\sinh^2(\mathcal{M}/2)}{\mathcal{\CM}^2}\,\nabla F\right]\!,
\end{equation}
where we have introduced the operators
\begin{equation}\label{eq:defofM}
 \nabla\,\cdot\ :=\ \d\cdot+\,[j_B,\,\cdot\,]\qquad\mbox{and}\qquad
 \mathcal{M}^2\,\cdot\,\ :=\ [F,[F,\,\cdot\,]].
\end{equation}
Note that in \eqref{eq:unfixedcurrent}, the first
term on the right hand side is proportional to the fermionic generators
of the superconformal algebra, while the second one is proportional to the bosonic ones.

\subsection{\texorpdfstring{$\kappa$}{Kappa}-symmetry gauge fixing}\label{sec:KRP}

Having fixed the local $H$-symmetry gauge in \rf{eq:rep1}, let us now discuss
a specific $\kappa$-symmetry gauge choice \cite{Kallosh:1998nx,Pesando:1998fv,Metsaev:2000yf,Metsaev:2000yu} that will  simplify the structure of the
string action and is natural in the present context. In the notation used in \rf{eq:rep1},
this gauge amounts to setting
\begin{equation}\label{sg}
 \theta^{i\alpha}_-\ =\ 0\ =\ \bar\theta^{\dot\alpha}_{-i},
\end{equation}
so that  the fermionic part  $\eu^{-F}$ of $g$  determined by
\begin{equation} \la{krg}
F(\Theta) \ =\  \im\big[ C_{ij}\epsilon_{\alpha\beta} \theta^{i\alpha}_+Q^{j\beta}
-C^{ij}\epsilon_{\dot\alpha\dot\beta} \bar\theta_{+i}^{\dot\alpha}
  \bar Q_j^{\dot\beta}\big],
\end{equation}
does not contain terms with $S$-generators (for that reason we shall
follow \cite{Metsaev:2000yf,Metsaev:2000yu} and refer to this gauge  as ``S-gauge'').

One may then readily check that in this case $\mathcal{M}^2\,\nabla F=0$. From 
Eq.~\eqref{eq:unfixedcurrent}, we deduce
\begin{subequations}
\begin{eqnarray}\label{eq:fercur-1}
 j_F &=& -\nabla F-\tfrac{1}{2}[F,\nabla F]\notag\\
     &=& -\im C_{ij}\epsilon_{\alpha\beta}\nabla\theta^{i\alpha}_+Q^{j\beta}
          +\im C^{ij}\epsilon_{\dot\alpha\dot\beta}\nabla\bar \theta^{\dot\alpha}_{+i}\bar Q^{\dot\beta}_j
 -\tfrac{1}{2}(\bar\theta^{\dot\alpha}_{+i}\nabla\theta^{i\beta}_+-\nabla\bar
   \theta^{\dot\alpha}_{+i}\theta^{i\beta}_+)
         P_{\beta\dot\alpha},
\end{eqnarray}
where
\begin{equation}\label{eq:fercur-2}
 \nabla\theta^{i\alpha}_+\ =\ \d\theta^{i\alpha}_++{\omega^{i\alpha}}_{j\beta}\theta^{j\beta}_+,\qquad
 \mbox{with}\qquad
 {\omega^{i\alpha}}_{j\beta}\ :=\ \tfrac{\im}{2}{\delta^\alpha}_\beta\big({\delta^i}_j j_{B_D}-
  C^{ki}C_{jl}j_{B_{{R_l}^k}}\big).
\end{equation}
\end{subequations}
Note that $\nabla\theta^{i\alpha}_+=(\nabla\bar\theta^{\dot\alpha}_{+i})^\dagger$.
Upon substituting the expressions of $j_{B_D}$ and $j_{B_{{R_i}^j}}$ given in \eqref{eq:boscur} into
\eqref{eq:fercur-2}, one may recast $\nabla\theta^{i\alpha}_+$ as
\begin{equation}
 \nabla\theta^{i\alpha}_+\ =\ Y^{1/2}{\Lambda^i}_k\,\d(Y^{-1/2}{(\Lambda^{-1})^k}_j\theta^{j\alpha}_+),
\end{equation}
where ${\Lambda^i}_j$ was defined  in \eqref{eq:lambda}. This then suggests performing
the following fermionic field redefinition:
\begin{equation}\la{ne}
 (\theta^{i\alpha}_+,\bar\theta^{\dot\alpha}_{+i})
\ \mapsto\ (\theta^{i\alpha},\bar\theta^{\dot\alpha}_i),\qquad
 \theta^{i\alpha}\ :=\ Y^{-1/2} {(\Lambda^{-1})^i}_j\theta^{j\alpha}_+, \qquad\mbox{with }\qquad
 \theta^{i\alpha}\ =\ (\bar \theta^{\dot\alpha}_i)^\dagger.
\end{equation}
Then the current $j=j_B+j_F$ expressed in terms
of the bosonic coordinates $(X,Y)$ and the new 16 independent
fermionic coordinates $\theta$ takes the following form
\begin{equation}\label{eq:fullcurrent}
\begin{aligned}
 j\ &=\ \im Y\big[ \d
 X^{\dot\alpha\beta}+\tfrac{\im}{2}(\bar\theta^{\dot\alpha}_i\d\theta^{i\beta}-\d\bar\theta^{\dot\alpha}_i
 \theta^{i\beta})\big] P_{\beta\dot\alpha}+\tfrac{\im}{Y}\d Y D +
   2\im {(\Lambda^{-1})^i}_k\d{\Lambda^k}_j{R_i}^j\ \\
  &\kern1.5cm -\ \im C_{ij}\epsilon_{\alpha\beta}Y^{1/2} {\Lambda^i}_k\d\theta^{k\alpha}Q^{j\beta}
          +\im C^{ij}\epsilon_{\dot\alpha\dot\beta}
                Y^{1/2} {(\Lambda^{-1})^k}_i\,\d\bar\theta^{\dot\alpha}_{k}\bar Q^{\dot\beta}_j.
\end{aligned}
\end{equation}
A particular choice of the coset representative and a particular
$\kappa$-symmetry gauge fixing  makes  part of the global symmetries non-manifest;
the manifest symmetries left after our gauge choices  are  the Poincar\'e translations,
Lorentz rotations, dilatations, $SU(4)$-rotations and the Poincar\'e (or $Q$, $\bar Q$) supersymmetry.

In order to write down the action \eqref{eq:action1} in the S-gauge,
we first need to extract the $j_{(m)}$-parts ($m=1,2,3$)
from the current \eqref{eq:fullcurrent}
using Eqs.~\eqref{eq:explicitsplitting}. We find (`h.c.' stands for Hermitian conjugation)
\begin{equation}\label{eq:decfullcurrent}
 \begin{aligned}
  j_{(1)}\ &=\ -\tfrac{\im}{2}C_{ij}\epsilon_{\alpha\beta}Y^{1/2} {\Lambda^i}_k\d\theta^{k\alpha}
               (Q^{j\beta}+C^{jl}S^\beta_l)\ -\ \mbox{h.c.},\\
  j_{(2)}\ &=\ \tfrac{\im}{2} Y\Pi^{\dot\alpha\beta}(P_{\beta\dot\alpha}+K_{\beta\dot\alpha})+
                     \tfrac{\im}{Y}\d Y D -  2\im C^{l[i}{(\Lambda^{-1})^{j]}}_k\d{\Lambda^k}_lR_{[ij]},\\
  j_{(3)}\ &=\ -\tfrac{\im}{2}C_{ij}\epsilon_{\alpha\beta}Y^{1/2} {\Lambda^i}_k\d\theta^{k\alpha}
               (Q^{j\beta}-C^{jl}S^\beta_l)\ -\ \mbox{h.c.},
 \end{aligned}
\end{equation}
where we defined
\begin{equation}\label{eq:superform}
 \Pi^{\dot\alpha\beta}\ :=\ \d
 X^{\dot\alpha\beta}+\tfrac{\im}{2}(\bar\theta^{\dot\alpha}_i\d\theta^{i\beta}-\d\bar\theta^{\dot\alpha}_i
 \theta^{i\beta}).
\end{equation}
Upon inserting the above expressions into the action \eqref{eq:action1}, we obtain
\cite{Kallosh:1998ji}
\begin{equation}\label{kta}
\begin{aligned}
 S\ &=\ -\tfrac{T}{2}\int_\Sigma\left\{-\tfrac{1}{2}
         Z^2\Pi_{\alpha\dot\beta}\wedge{*\Pi}^{\dot\beta\alpha}
        +\tfrac{1}{4 Z^2}\,\d Z_{ij}\wedge{*\d}Z^{ij}\ \right.\\
    &\left.\kern2cm+\ \tfrac{1}{2}
     \bigl(\epsilon_{\alpha\beta}\,\d Z_{ij}\wedge\theta^{i\alpha}\d\theta^{j\beta}
    -\epsilon_{\dot\alpha\dot\beta}\,\d Z^{ij}\wedge\bar\theta^{\dot\alpha}_i\d\bar\theta^{\dot\beta}_j\bigr)
\right\}\!.
\end{aligned}
\end{equation}
In deriving this form of the action, we have used the invariant form on $\mathfrak{psu}(2,2|4)$ given in
\eqref{eq:cartan-killing} and performed the change of coordinates $Y_{ij}\mapsto Z_{ij}$, with
\begin{equation}\label{tildeY}
 Z_{ij}\ :=\ YC_{kl}{\Lambda^k}_i{\Lambda^l}_j,\qquad
 Z_{ik} Z^{jk}\ =\ Y^2{\delta_i}^j\qquad\mbox{and}\qquad Z^2\ =\ \tfrac{1}{4}Z_{ij}Z^{ij}\ =\ Y^2.
\end{equation}
As a result, the action does not depend on the choice of the constant matrix $C_{ij}$.

\subsection{Gauge fixed  form of flat currents}
Inserting the expressions \eqref{eq:decfullcurrent} into
\eqref{eq:1PF}, we immediately arrive at the $\kappa$-gauge
fixed version of the family of flat currents  $\hj(z)$. The construction of the
other family of flat currents $\hJ(z)$ given in
\eqref{eq:ginvflatcur} requires more work.
Firstly, notice that we might rewrite \eqref{eq:ginvflatcur} as
\begin{subequations}
\begin{equation}
  \hJ(z)\ =\ (z-1)J_{(1)}+\tfrac{1}{2}(z^2+z^{-2}-2)J_{(2)}+
                 (z^{-1}-1)J_{(3)}-\tfrac{1}{2}(z^2-z^{-2}){*J_{(2)}},
\end{equation}
where we have defined
\begin{equation}
  J_{(m)}\ :=\ gj_{(m)}g^{-1}.
\end{equation}
\end{subequations}
Upon using successively the
Baker-Campbell-Hausdorff formula, we arrive after some rather
lengthy algebraic manipulations at
\begin{subequations}\label{eq:NCpoincare}
\begin{eqnarray}
 J_{(1)} &=&  -\tfrac{1}{2}\bigl(
               \bar\theta^{\dot\alpha}_i\d\theta^{i\beta}-
                   Z_{ij} X^{\dot\alpha\gamma}\d\theta^i_\gamma\theta^{j\beta}\bigr)P_{\beta\dot\alpha}
             +\tfrac{1}{3!}\epsilon_{\gamma\delta}
               \d\theta^j_\alpha\bar\theta_{k\dot\beta}\bigl(\theta^{j\delta}\theta^{k(\alpha}P^{\dot\beta\gamma)}
               +\theta^{i(\alpha}\theta^{k\delta)}P^{\dot\beta\gamma}\bigr) \notag\\[4pt]
            &&\kern.9cm   +\ \tfrac{\im}{2}Z_{ij}\d\theta^{i\alpha}\theta^{j\beta}L_{\alpha\beta}
               + \tfrac{\im}{4}Z_{ij}\d\theta^{i\alpha}\theta^j_\alpha D
               + \im Z_{ij}\d\theta^{k\alpha}\theta^i_\alpha {R_k}^j
                \notag\\[4pt]
            &&\kern1.2cm   +\ \tfrac{\im}{ Z} Z_{ij}\d\theta^i_\alpha Q^{j\alpha}-
               \tfrac{\im}{2 Z^2} Z_{ij} Z_{kl}\epsilon_{\beta\gamma}\d\theta^i_\alpha
                    \bigl(\theta^{j(\alpha}\theta^{k\gamma)}Q^{l\beta}+\tfrac{1}{2}
                      \theta^{j\gamma}\theta^{k\beta}Q^{l\alpha}\bigr)\notag\\[4pt]
            && \kern1.5cm   +\ \tfrac{1}{2}
               \epsilon_{\dot\alpha\dot\beta}\d\theta^i_\gamma
                      \bigl(X^{\dot\alpha\gamma}-\tfrac{\im}{2}\bar\theta^{\dot\alpha}_j\theta^{j\gamma}\bigr)
                      \bar Q^{\dot\beta}_i+\tfrac{\im}{2} Z\d\theta^i_\alpha S^\alpha_i\,
                \, -\, \mbox{h.c.},\\[4pt]
 J_{(2)} &=& \left\{\im  Z\Pi^{\dot\delta\gamma}
                  A_{\gamma\dot\delta}^{\dot\alpha\beta}+\tfrac{\im}{ Z}\d
                 Z\bigl(X^{\dot\alpha\beta}+\tfrac{\im}{2}
              \bar\theta^{\dot\alpha}_i\theta^{i\beta}\bigr)-\tfrac{1}{2 Z^2} Z^{ik}\d
                 Z_{kj}\bar\theta^{\dot\alpha}_i \theta^{j\beta}\right\}P_{\beta\dot\alpha}\, \notag\\
                &&~~-\, \tfrac{\im}{2} Z^2
                \Pi_{\alpha\dot\beta}\epsilon_{\gamma\delta}\bigl(X^{\dot\beta\delta}-
                     \tfrac{\im}{2}\bar\theta^{\dot\beta}_i\theta^{i\delta}\bigl)L^{\alpha\gamma}
                         -\tfrac{\im}{2} Z^2
                      \Pi_{\beta\dot\alpha}\epsilon_{\dot\gamma\dot\delta}\bigl(X^{\dot\delta\beta}+
                     \tfrac{\im}{2}\bar\theta^{\dot\delta}_i\theta^{i\beta}\bigl)L^{\dot\alpha\dot\gamma}\,
                      \notag\\[4pt]
              &&~~~+\, \bigl(\tfrac{\im}{2} Z^2\Pi_{\beta\dot\alpha}X^{\dot\alpha\beta}
                          +\tfrac{\im}{ Z}\d Z\bigr)D+\tfrac{\im}{2} Z^2
                     \Pi^{\dot\alpha\beta}K_{\beta\dot\alpha}\notag\\[4pt]
              &&~~~~+\, \bigl(\Pi_{\beta\dot\alpha} Z^{ik} Z_{jl}\bar\theta^{\dot\alpha}_k
                   \theta^{l\beta}-\tfrac{\im}{ Z^2} Z^{ik}\d Z_{kj}\bigr){R_i}^j  \notag\\[4pt]
              &&~~~~~+\, \im\bigl( Z\Pi_{\beta\dot\alpha} B^{\dot\alpha\beta}_{i\gamma}-
                       \tfrac{1}{2 Z}\d Z_{ij}\theta^j_\gamma\bigr)Q^{i\gamma}-
                    \im\bigl( Z\Pi_{\alpha\dot\beta} (B^{\dot\alpha\beta}_{i\gamma})^\dagger-
                       \tfrac{1}{2 Z}\d Z^{ij}\bar\theta_{j\dot\gamma}\bigr)\bar Q_i^{\dot\gamma}
                       \, \notag\\[4pt]
                &&~~~~~~-\,\tfrac{1}{2} Z Z^{ij}\Pi_{\alpha\dot\beta}\bar\theta^{\dot\beta}_jS^\alpha_i
                - \tfrac{1}{2} Z Z_{ij}\Pi_{\beta\dot\alpha}\theta^{j\beta}\bar S^{i\dot\alpha},\\[4pt]
  J_{(3)} &=&  \left\{\tfrac{\im}{ Z} Z_{ij}\d\theta^i_\alpha Q^{j\alpha}
                   -\bar
           \theta^{\dot\alpha}_i\d\theta^{i\beta}P_{\beta\dot\alpha}\, -\, \mbox{h.c.}\right\}-J_{(1)}.
\end{eqnarray}
Here, we have defined
\begin{eqnarray}
 A^{\dot\alpha\beta}_{\gamma\dot\delta} &:=& \tfrac{1}{2 Z}\bigl(1+X^2 Z^2\bigr)
          {\delta_{\dot\delta}}^{\dot\alpha}{\delta_\gamma}^\beta+\tfrac{1}{2} ZX^{\dot\alpha\beta}
          X_{\gamma\dot\delta}\notag\\
         &&\kern.5cm-\ \tfrac{\im}{4} Z\left(\epsilon^{\dot\alpha\dot\beta}
            X_{\alpha\dot\beta}\bar\theta_{i\dot\delta}\theta^{i(\alpha}{\delta_\gamma}^{\beta)}
               +\tfrac{1}{3! Z^2} Z^{ij}\bar\theta_{i\dot\delta}B^{\dot\alpha\beta}_{j\gamma}
                \ -\ \mbox{h.c.}\right)\!,\\
 B^{\dot\alpha\beta}_{i\gamma} &:=& -\tfrac{\im}{ Z} Z_{ij}\left(\bar\theta^{\dot\alpha}_k\theta^k_\delta
               \theta^{j(\delta}\delta_\gamma^{\beta)}+2\delta_\gamma^\delta
                \bigl(\bar\theta^{\dot\alpha}_k\theta^k_\delta\theta^{j\beta}-\tfrac{1}{2}
                 \theta^j_\delta\theta^{k\beta}\bar\theta^{\dot\alpha}_k\bigr)\right)\!,
\end{eqnarray}
\end{subequations}
with $X^2:=-\frac12 X_{\alpha\dot\beta}X^{\dot\beta\alpha}$.
Note that we have again  performed the change of coordinates  \eqref{tildeY}.
Also note that in the final expression
of the currents, the $Sp(4)$-metric $C_{ij}$ does not appear as expected
in view of  $SU(4)$-invariance.

If we set $\theta^{i\alpha}=0=\bar\theta^{\dot\alpha}_i$, the fermionic parts
$J_{(1)}$ and $J_{(3)}$ become  identically zero, while the bosonic part
$J_{(2)}$ reduces to
\begin{eqnarray}\label{eq:bosnicnoether}
 J_{(2)} &=&  \tfrac{\im}{2}\left\{\bigl(1+X^2 Z^2\bigr)
                \d X_{\alpha\dot\beta}+ Z^2X^{\dot\gamma\delta}
                \d X_{\delta\dot\gamma} X_{\alpha\dot\beta}+
                \tfrac{2}{ Z}\d ZX_{\alpha\dot\beta}\right\}P^{\dot\beta\alpha}\,\notag\\
           &&\kern.7cm-\,\tfrac{\im}{2} Z^2
                \d X_{\alpha\dot\beta}\epsilon_{\gamma\delta}X^{\dot\beta\delta}L^{\alpha\gamma}
                     -\tfrac{\im}{2} Z^2
               \d X_{\beta\dot\alpha}\epsilon_{\dot\gamma\dot\delta}X^{\dot\delta\beta}
                     L^{\dot\alpha\dot\gamma}\,\notag\\[3pt]
             &&\kern1.2cm+\,\bigl(\tfrac{\im}{2}
                      Z^2\d X_{\alpha\dot\beta}X^{\dot\beta\alpha}
                          +\tfrac{\im}{ Z}\d Z\bigr)D+\tfrac{\im}{2} Z^2
                     \d X_{\alpha\dot\beta}K^{\dot\beta\alpha}-
               \tfrac{\im}{ Z^2} Z^{ik}\d Z_{kj}{R_i}^j.
\end{eqnarray}
These are precisely the Noether currents for the bosonic sigma model on AdS$_5\times S^5$ in the metric
\eqref{eq:ini}, i.e.\ the Noether currents associated with the Killing vectors of \eqref{eq:ini}
(see also \cite{Ricci:2007eq}).

\

\section{Duality transformation on \texorpdfstring{${\rm AdS}_5$}{AdS5} coordinates (bosonic T-duality) }\label{sec:BTduality}

Let us now turn to the discussion of the duality transformation of the superstring sigma model
along the four isometry directions $X^{\dot\alpha\beta}$ of AdS$_5$ in the Poincar\'e
coordinates following \cite{Kallosh:1998ji}.\foot{Note that we are performing the duality along the 
non-compact directions, i.e.\ as in \cite{Kallosh:1998ji}, we are concerned here with a formal sigma model duality.
We shall still refer to this as T-duality. Let us mention that T-duality transformations of
type II superstrings were discussed also, e.g., in \cite{Cvetic:1999zs,Kulik:2000nr,Hassan:2000kr,Alday:2005ww}.} In particular, we will generalize
the results of \cite{Ricci:2007eq} and explain how to construct families of flat currents for the T-dual model,
making  its integrability manifest.

\subsection{T-duality transformation of the superstring action}

To implement the  duality along $X$,  let us  start with the
first-order form of the action \eqref{kta} (see also \cite{Buscher:1987qj,Rocek:1991ps})
\begin{equation}\label{eq:roc}
\begin{aligned}
 S\ &=\ -\tfrac{T}{2}\int_\Sigma\left\{-\tfrac{1}{2}
          Z^2 \bigl( {\rm V}^{\dot\alpha\beta}
         +\tfrac{\im}{2}(\bar\theta^{\dot\alpha}_i\d\theta^{i\beta}-\d\bar\theta^{\dot\alpha}_i
 \theta^{i\beta})\bigr)\wedge{*\bigl( {\rm V}_{\beta\dot\alpha}+
 \tfrac{\im}{2}(\theta^i_\beta\d\bar\theta_{i\dot\alpha}-\d\theta^i_\beta\bar\theta_{i\dot\alpha}
 )\bigr)}\ \right.\\
      &~~~~~~~~~~~~~\left.+\ \tilde X_{\alpha\dot\beta}
         \d  {\rm V}^{\dot\beta\alpha}
        +\tfrac{1}{4 Z^2}\,\d  Z_{ij}\wedge{*\d} Z^{ij}
    + \tfrac{1}{2}
     \bigl(\d  Z_{ij}\wedge\theta^{i\alpha}\d\theta^j_\alpha
    -\d  Z^{ij}\wedge\bar\theta^{\dot\alpha}_i\d\bar\theta_{j\dot\alpha}\bigr)
\right\}.
\end{aligned}
\end{equation}
where  $ {\rm V} $ is an auxiliary one-form field and the field $\tilde X^{\alpha\dot\beta}$
(which will become the T-dual coordinate) plays the role of a Lagrange multiplier imposing
the flatness of ${\rm V}$, i.e.\ $\d{\rm V}=0 \Rightarrow {\rm V} = \d X$. On the other hand, solving for 
${\rm V}$ first yields
\begin{equation}
 {\rm V}^{\dot\alpha\beta}+\tfrac{\im}{2}(\bar\theta^{\dot\alpha}_i\d\theta^{i\beta}-
 \d\bar\theta^{\dot\alpha}_i \theta^{i\beta})\ =\  Z^{-2}\,{*\d}\tilde X^{\dot\alpha\beta},
\end{equation}
and thus the T-dual action written in terms of $\tilde X^{\dot\beta\alpha}$ becomes
\cite{Kallosh:1998ji}
\begin{equation}\label{dukt}
\begin{aligned}
 S\ &=\ -\tfrac{T}{2}\int_\Sigma\Big\{-\tfrac{1}{2 Z^2}
         \d\tilde X_{\alpha\dot\beta}\wedge{*\d\tilde X}^{\dot\beta\alpha}
        +\tfrac{1}{4 Z^2}\,\d  Z_{ij}\wedge{*\d} Z^{ij}\ \\
    &\kern2.5cm+\ \tfrac{\im}{2} \d\tilde X_{\beta\dot\alpha}\wedge
      (\bar\theta^{\dot\alpha}_i\d\theta^{i\beta}-\d\bar\theta^{\dot\alpha}_i\theta^{i\beta})+ \tfrac{1}{2}
     \bigl(\d  Z_{ij}\wedge\theta^{i\alpha}\d\theta^j_\alpha
    -\d  Z^{ij}\wedge\bar\theta^{\dot\alpha}_i\d\bar\theta_{j\dot\alpha}\bigr)
\Big\}.
\end{aligned}
\end{equation}
One observes that the bosonic geometry is again AdS$_5\times S^5$ (to put the bosonic
action into  the exactly same form one needs to change  coordinates $Z_{ij}$ so that
$Z \mapsto Z^{-1}$). Also, the dual action is quadratic in the fermions.
Moreover, the fermionic part of the action
is of WZ type and therefore does not depend on the worldsheet metric.\footnote{\label{foot:shift}Note that
when integrating out ${\rm V}_{\alpha\dot\beta}$ in the path integral, one picks up a
 factor $\Delta_B^{-1/2}$
involving the functional determinant $\Delta_B=\Pi_{\sigma\in\Sigma} Z^{16}(\sigma)$ 
(in units where $T=-2$) which needs to be regularized. Using heath kernel methods, this
amounts to adding the term $-8\int_\Sigma{\rm dvol}\,R^{(2)}\log(Z)$
to the action \eqref{dukt} (cf.~\cite{Buscher:1987qj,Schwarz:1992te}),
$R^{(2)}$ being the scalar curvature of $\Sigma$.} 

Let us remark that the on-shell relation between the original and dual coordinates is
\begin{equation}\label{dtr}
 \d X^{\dot\alpha\beta}+\tfrac{\im}{2}(\bar\theta^{\dot\alpha}_i\d\theta^{i\beta}-
 \d\bar\theta^{\dot\alpha}_i
 \theta^{i\beta})\ =\   Z^{-2}\,{*\d}\tilde X^{\dot\alpha\beta}.
\end{equation}

\subsection{Flat currents for the T-dual model}

In general, if the original model is classically integrable, the same  applies to its
dual counterpart: the flatness of the Lax connection gives first-order equations that
``interpolate'' between the original and dual model. Still, it is useful to find the explicit
expression for the flat currents in terms of the dual coordinates as this may also help clarify
the transformation of the conserved charges under the T-duality.

For our choice of the Poincar\'e coordinates
 and the $\kappa$-symmetry gauge, the current $j$ depends on the
 original coordinate $X^{\dot\alpha\beta}$
only through its  differential $\d X^{\dot\alpha\beta}$, see
Eq.~\eqref{eq:fullcurrent}. The same then
applies  to the  family of flat currents $\hj(z)$  in \rf{eq:1PF},
where the expressions for $j_{(1)}$, $j_{(2)}$ and $j_{(3)}$ are given in
\eqref{eq:decfullcurrent} and
\begin{equation}
 A\ =\ j_{(0)}\ =\ \tfrac{\im}{2} Y\Pi^{\dot\alpha\beta}(P_{\beta\dot\alpha}-K_{\beta\dot\alpha})
                    -  2\im C^{l(i}{(\Lambda^{-1})^{j)}}_k\d{\Lambda^k}_lR_{(ij)}.
\end{equation}
Then it is straightforward to re-express  $\hj(z)$ in terms of $\td X$ by
using \rf{dtr}, i.e.\ by replacing $\Pi_{\alpha\dot\beta}$ with
$Z^{-2}\,{*\d}\tilde X_{\alpha\dot\beta}$. The resulting family of currents 
$\tilde \hj := \hj (X \mapsto \td X)$ is still flat since \eqref{dtr} holds
on-shell. And vice versa, the flatness of $\tilde \hj$ will imply the field 
equations of the T-dual model.

As already discussed above, the family
$\hj(z)$ is not $H$-gauge invariant, i.e.\ it depends on a choice of representative of $G/H$.
In order to  be able to discuss  the physical conserved  charges, it is therefore
useful to repeat the same procedure of replacing $X$ by $\td X$
for the other family of flat currents $\hJ(z)$ in \rf{eq:ginvflatcur}
closely related to Noether charge.
However, unlike $\hj(z)$, the current  $\hJ(z)$
which involves  $g$    explicitly depends
on $X$  and thus, if dualized directly, would
  non-locally depend on $\td X$.
One can by-pass this problem and get a local expression for $\hJ(z)$
in terms of the dual coordinate  $\td X$ by first performing
a $G$-gauge transformation \rf{gah} (which preserves the flatness condition \rf{eq:flat2})
with the following parameter ${\rm g}$:
\begin{equation}
 {\rm g}\ =\ \eu^{\im X^{\dot\alpha\beta}P_{\beta\dot\alpha}}.
\end{equation}
Then the gauge transformed current is
\begin{equation}
  \hJ'(z)\ =\ (z-1)J'_{(1)}+\tfrac{1}{2}(z^2+z^{-2}-2)J'_{(2)}+
                 (z^{-1}-1)J'_{(3)}-\tfrac{1}{2}(z^2-z^{-2}){*J'_{(2)}}
      +\underbrace{\im \d X^{\dot\alpha\beta}P_{\beta\dot\alpha}}_{=\,{\rm g}^{-1}\d{\rm g}},
\end{equation}
with
\begin{equation}
  J'_{(m)}\ :=\ {\rm g}^{-1}J_{(m)}{\rm g}\ =\ g' j_{(m)} g'^{-1},\qquad\mbox{with}\qquad g'={\rm g}^{-1}g,
\end{equation}
where  $g$ is given  by  \eqref{eq:rep1} in the  S-gauge \eqref{sg}.\foot{Note that
 $\hJ'(z)$ is invariant under the  $H$-gauge transformations \eqref{eq:htrafo}:
 under such transformations
${\rm g}\mapsto{\rm g}$ and $g'\mapsto g'h$ and thus
$g'j_{(m)}g'^{-1}\mapsto g'h(h^{-1}j_{(m)}h) h^{-1} g'^{-1}=g'j_{(m)}g'^{-1}$
and ${\rm g}^{-1}\d{\rm g}\mapsto{\rm g}^{-1}\d{\rm g}$.}
Then  $g'=g(X=0,Y,\theta_+,\theta_-=0)$ and thus (cf.~Eqs.~\eqref{eq:NCpoincare})
\begin{subequations}
\begin{eqnarray}
 J'_{(1)} &=&  -\tfrac{1}{2}
               \bar\theta^{\dot\alpha}_i\d\theta^{i\beta}P_{\beta\dot\alpha}
             +\tfrac{1}{3!}\epsilon_{\gamma\delta}
               \d\theta^j_\alpha
           \bar\theta_{k\dot\beta}\bigl(\theta^{j\delta}\theta^{k(\alpha}P^{\dot\beta\gamma)}
               +\theta^{i(\alpha}\theta^{k\delta)}P^{\dot\beta\gamma}\bigr) \notag\\[4pt]
            &&\kern.9cm   +\ \tfrac{\im}{2} Z_{ij}\d\theta^{i\alpha}\theta^{j\beta}L_{\alpha\beta}
               + \tfrac{\im}{4} Z_{ij}\d\theta^{i\alpha}\theta^j_\alpha D
               + \im Z_{ij}\d\theta^{k\alpha}\theta^i_\alpha {R_k}^j
                \notag\\[4pt]
            &&\kern1.2cm   +\ \tfrac{\im}{ Z} Z_{ij}\d\theta^i_\alpha Q^{j\alpha}-
               \tfrac{\im}{2 Z^2} Z_{ij} Z_{kl}\epsilon_{\beta\gamma}\d\theta^i_\alpha
                    \bigl(\theta^{j(\alpha}\theta^{k\gamma)}Q^{l\beta}+\tfrac{1}{2}
                      \theta^{j\gamma}\theta^{k\beta}Q^{l\alpha}\bigr)\notag\\[4pt]
            && \kern1.5cm   -\ \tfrac{\im}{4}
               \epsilon_{\dot\alpha\dot\beta}\d\theta^i_\gamma
                      \bar\theta^{\dot\alpha}_j\theta^{j\gamma}
                      \bar Q^{\dot\beta}_i+\tfrac{\im}{2} Z\d\theta^i_\alpha S^\alpha_i\,
                \, -\, \mbox{h.c.},
\end{eqnarray}
\begin{eqnarray}
 J'_{(2)} &=& \left\{\im  Z\Pi^{\dot\delta\gamma}
                  A_{\gamma\dot\delta}^{\dot\alpha\beta}-\tfrac{1}{2 Z}\d
                 Z
              \bar\theta^{\dot\alpha}_i\theta^{i\beta}-\tfrac{1}{2 Z^2} Z^{ik}\d
                 Z_{kj}\bar\theta^{\dot\alpha}_i \theta^{j\beta}\right\}P_{\beta\dot\alpha}\, \notag\\
                &&~~-\, \tfrac{1}{4} Z^2
                \Pi_{\alpha\dot\beta}\epsilon_{\gamma\delta}
                 \bar\theta^{\dot\beta}_i\theta^{i\delta}L^{\alpha\gamma}
                         +\tfrac{1}{4} Z^2
                      \Pi_{\beta\dot\alpha}\epsilon_{\dot\gamma\dot\delta}
                     \bar\theta^{\dot\delta}_i\theta^{i\beta}L^{\dot\alpha\dot\gamma}\,
                      \notag\\[4pt]
              &&~~~+\,\tfrac{\im}{ Z}\d ZD+\tfrac{\im}{2} Z^2
                     \Pi^{\dot\alpha\beta}K_{\beta\dot\alpha}
              + \bigl(\Pi_{\beta\dot\alpha} Z^{ik} Z_{jl}\bar\theta^{\dot\alpha}_k
                   \theta^{l\beta}-\tfrac{\im}{ Z^2} Z^{ik}\d Z_{kj}\bigr){R_i}^j\notag\\[4pt]
              &&~~~~+\, \im\bigl( Z\Pi_{\beta\dot\alpha} B^{\dot\alpha\beta}_{i\gamma}-
                       \tfrac{1}{2 Z}\d Z_{ij}\theta^j_\gamma\bigr)Q^{i\gamma}-
                    \im\bigl( Z\Pi_{\alpha\dot\beta} (B^{\dot\alpha\beta}_{i\gamma})^\dagger-
                       \tfrac{1}{2 Z}\d Z^{ij}\bar\theta_{j\dot\gamma}\bigr)\bar Q_i^{\dot\gamma}
                       \, \notag\\[4pt]
                &&~~~~~-\,\tfrac{1}{2} Z Z^{ij}\Pi_{\alpha\dot\beta}\bar\theta^{\dot\beta}_jS^\alpha_i
                - \tfrac{1}{2} Z Z_{ij}\Pi_{\beta\dot\alpha}\theta^{j\beta}\bar S^{i\dot\alpha},\\[4pt]
  J'_{(3)} &=&  \left\{\tfrac{\im}{ Z} Z_{ij}\d\theta^i_\alpha Q^{j\alpha}
                   -\bar \theta^{\dot\alpha}_i\d\theta^{i\beta}P_{\beta\dot\alpha}\, -\,
            \mbox{h.c.}\right\}-J'_{(1)},
\end{eqnarray}
where here
\begin{eqnarray}
 A^{\dot\alpha\beta}_{\gamma\dot\delta} &:=& \tfrac{1}{2 Z}
          {\delta_{\dot\delta}}^{\dot\alpha}{\delta_\gamma}^\beta
        - \tfrac{\im}{4! Z}\left(
               Z^{ij}\bar\theta_{i\dot\delta}B^{\dot\alpha\beta}_{j\gamma}
                \ -\ \mbox{h.c.}\right)\!,\\
 B^{\dot\alpha\beta}_{i\gamma} &:=& -\tfrac{\im}{ Z} Z_{ij}\left(\bar\theta^{\dot\alpha}_k\theta^k_\delta
               \theta^{j(\delta}\delta_\gamma^{\beta)}+2\delta_\gamma^\delta
                \bigl(\bar\theta^{\dot\alpha}_k\theta^k_\delta\theta^{j\beta}-\tfrac{1}{2}
                 \theta^j_\delta\theta^{k\beta}\bar\theta^{\dot\alpha}_k\bigr)\right)\!.
\end{eqnarray}
\end{subequations}
Note that the  bosonic truncation  of  the gauge transformed current $\hJ'(z)$ is given by
\begin{equation}
 \begin{aligned}
 J'(z)\ &=\ \tfrac{1}{2}(z^2+z^{-2}-2) J'_{(2)}+\tfrac{1}{2}(z^2-z^{-2})\, {*J'_{(2)}}+
    \im \d X^{\dot\alpha\beta}P_{\beta\dot\alpha}, \\
 J'_{(2)}\ &=\ \tfrac{\im}{2}(1+ Z^2)\d X_{\alpha\dot\beta}P_+^{\dot\beta\alpha}+
                \tfrac{\im}{2}(1- Z^2)\d X_{\alpha\dot\beta}P_-^{\dot\beta\alpha}
             +\tfrac{\im}{ Z}\d Z D-
               \tfrac{\im}{ Z^2} Z^{ik}\d Z_{kj}{R_i}^j,
 \end{aligned}
\end{equation}
where $P^{\dot\alpha\beta}_\pm:=\tfrac12(P^{\dot\alpha\beta}\pm
K^{\dot\alpha\beta})$. Up to a rotation by a constant matrix,
these are the same gauge transformed currents as found directly in Ref.~\cite{Ricci:2007eq}
without referring to the coset nature  of the AdS space.

Finally, using the duality relation  in Eq.~\eqref{dtr}, we find the expressions for
the currents in terms of the T-dual coordinates:
\begin{equation}
  \tilde \hJ(z;\tilde X, Z,\Theta)\ :=\ \hJ'(z;X(\tilde X), Z,\Theta), \la{lol}
\end{equation}
i.e.
\begin{subequations}
 \begin{equation}
  \begin{aligned}
  \tilde \hJ(z)\ &=\ (z-1)\tilde J_{(1)}+\tfrac{1}{2}(z^2+z^{-2}-2)\tilde J_{(2)}+
                 (z^{-1}-1)\tilde J_{(3)}-\tfrac{1}{2}(z^2-z^{-2}){*\tilde J_{(2)}}\  \\
    &\kern1.5cm+ \tfrac{\im}{ Z^2}\big[{*\d}\tilde X^{\dot\alpha\beta}-\tfrac{\im}{2} Z^2\bigl(
        \bar\theta^{\dot\alpha}_i\d\theta^{i\beta}-\d\bar\theta^{\dot\alpha}_i\theta^{i\beta}\bigr)\big]\!
       P_{\beta\dot\alpha},
  \end{aligned}
\end{equation}
with
\begin{equation}
 \tilde J_{(m)}\ :=\ J'_{(m)}\bigl(\Pi_{\alpha\dot\beta}\mapsto Z^{-2}{*\d}\tilde X_{\alpha\dot\beta}\bigr).
\end{equation}
\end{subequations}
Note that $\tilde J(z)$ is flat since the duality relation \eqref{dtr} holds on-shell.
Having  expressed the  Lax connection in terms of the dual coordinates, one can in
principle derive an infinite set of non-local charges in the T-dual model by using \eqref{charge}.

\subsection{Bosonic duality as a symmetry of first-order system and Lax connection}\label{BB}

With a motivation to eventually shed some light on how conserved
charges of  the original and  T-dual models are related, let us go
back to the purely bosonic sigma model on AdS$_5\cong
SO(2,4)/SO(1,4)$.\footnote{The subsequent discussions can of
course be applied to the AdS$_n$-case.} We shall ignore  the decoupled
5-sphere part here. As was shown in \cite{Ricci:2007eq}, the  T-duality
applied to the   bosonic AdS$_n$-model  generically maps
conserved local charges into non-local ones and vice versa.
To make  this more precise  it is desirable
to describe T-duality as a formal algebraic transformation on the phase space,
i.e.\ on the components of the  current  subject to first-order equations.

Let us begin by recalling the $\IZ_2$-automorphism of
the conformal $\mathfrak{g}=\mathfrak{so}(2,4)$ algebra (cf.~Appendix \ref{sec:SCA})
\begin{equation}\label{eq:z2auto}
\Omega\bigl(P_{\alpha\dot\beta}\bigr)\ =\ -K_{\alpha\dot\beta},\quad
\Omega\bigl(K_{\alpha\dot\beta}\bigr)\ =\ -P_{\alpha\dot\beta},\quad
 \Omega\left(L_{\alpha\beta}\right)\ =\ L_{\alpha\beta}\quad\mbox{and}\quad \Omega\left(D\right)\ =\ -D.
\end{equation}
Then we may define the projectors
\begin{equation}
 \cP_{(0)}\ :=\ \tfrac{1}{2}(1+ \Omega)\qquad\mbox{and}\qquad
 \cP_{(2)}\ :=\ \tfrac{1}{2}(1- \Omega),
\end{equation}
so that
\begin{equation}
 \mathfrak{g}\ \cong\ \mathfrak{h}\oplus\mathfrak{g}_{(2)},\qquad\mbox{with}\qquad
 \mathfrak{h}\ =\ \cP_{(0)}(\mathfrak{g})\qquad\mbox{and}\qquad\mathfrak{g}_{(2)}\ =\ \cP_{(2)}(\mathfrak{g}).
\end{equation}
Correspondingly, the current $j=g^{-1}\d g$, for  $g\in SO(2,4)$, decomposes as
\begin{equation}
  j\ =\ A+j_{(2)},\qquad\mbox{with}\qquad \cP_{(0)}(j)\ =\ A\qquad\mbox{and}\qquad \cP_{(2)}(j)\ =\ j_{(2)}.
\end{equation}
The Maurer--Cartan equations and the equations of motion are the same as in \rf{coo} or
 found  by setting $j_{(1)}=0=j_{(3)}$ in
\eqref{iden} and \eqref{eom}
\begin{equation}\label{eq:bosoniceq}
 \d A + A \wedge A +j_{(2)}\wedge j_{(2)}\ =\ 0,\qquad\nabla j_{(2)}\ =\ 0
 \qquad\mbox{and}\qquad\nabla {*j}_{(2)}\ =\ 0.
\end{equation}
For the  choice of the AdS$_5$-part of the  coset representative  in \eqref{eq:rep1},
i.e.\ $g= \eu^{\im X^{\dot\alpha\beta} P_{\beta\dot\alpha}} Y^{\im D}$,
we have
\begin{equation}\la{jap}
 j\ =\ j_P+j_D,\qquad\mbox{with}\qquad j_P\ =\ \im Y \d X^{\dot\alpha\beta} P_{\beta\dot\alpha}
 \qquad\mbox{and}\qquad j_D\ =\ \tfrac{\im}{Y} \d Y D.
\end{equation}
In this parametrization, Eqs.~\eqref{eq:bosoniceq} read explicitly as
\begin{equation}\label{phase}
\begin{aligned}
 \d j_P+j_D\wedge j_P+j_P\wedge j_D\ &=\ 0,\\
\d j_D\ &=\ 0,\\
\d {*j}_P-j_D\wedge {*j}_P-{*j}_P\wedge j_D\ &=\ 0, \\
\d {*j}_D-\tfrac{1}{2}j_P\wedge {*\Omega}\left(j_P\right)-\tfrac{1}{2}{*\Omega}\left(j_P\right)\wedge
j_P\ &=\ 0.
\end{aligned}
\end{equation}
Here, $\Omega(j_P)=\Omega(j_{_{P^{\dot\alpha\beta}}}P^{\dot\alpha\beta})=j_{_{P^{\dot\alpha\beta}}}
\Omega(P^{\dot\alpha\beta})=
-j_{_{P^{\dot\alpha\beta}}}K^{\dot\alpha\beta}$.
The  T-duality along the four isometry directions $X^{\dot\alpha\beta}$ corresponds to replacing
$(X^{\dot\alpha\beta},Y)$ by the dual fields $(\tilde X^{\dot\alpha\beta},\tilde Y)$
according to\footnote{Here, the field $Y$ appears instead of $Z$ when
compared with \eqref{dtr} since we dropped the
$S^5$-part and so  the field redefinition \eqref{tildeY} is not needed.}
\begin{equation}
 \d\tilde X^{\dot\alpha\beta}\ =\ Y^2 {*\d}X^{\dot\alpha\beta}\qquad\mbox{and}\qquad
 \tilde Y\ =\ Y^{-1}.
\end{equation}
Therefore, the components of the dual current
$\tilde j=\tilde j_P+\tilde j_D$  is defined in terms of
 $(\tilde X^{\dot\alpha\beta},\tilde Y)$
in  exactly the same way as $j$ in \rf{ja}  is defined in terms of
$(X^{\dot\alpha\beta}, Y)$. It
can then  be expressed in terms of the original coordinates as follows:
\begin{equation}
\begin{aligned}
 j_P\ &=\ \im Y \d  X^{\dot\alpha\beta} P_{\beta\dot\alpha}\ =\ \im \tilde Y {*\d}
\tilde X^{\dot\alpha\beta}P_{\beta\dot\alpha}\ =\ {*\tilde j}_P,\\
 j_D\ &=\  \tfrac{\im}{Y} \d Y D\ =\ -\tfrac{\im}{\tilde Y}\d \tilde YD\ =\ -\tilde j_D.
\end{aligned}
\end{equation}
The key point is that under  this  transformation, i.e.
\begin{equation}\label{dul}
 j_P\ \mapsto\ \tilde j_P\ =\ {*j}_P\qquad\mbox{and}\qquad j_D\ \mapsto\ \tilde j_D\ =\ -j_D,
\end{equation}
the set of first-order equations \eqref{phase} is invariant;  in particular,
the Maurer--Cartan equation for $j_P$ is interchanged with its equation of motion.
Thus, we may forget  about particular solutions for $j$ in terms of $X$ or $\tilde X$ and view
the duality as a symmetry of the phase space equations \rf{phase}.

The family of flat currents \eqref{eq:1PF} here takes the form
\begin{equation}\label{eq:lax1}
\begin{aligned}
\hj(z)\ &=\ \tfrac{1}{4}(z+z^{-1})^2 j_P-\tfrac{1}{4}(z-z^{-1})^2 \Omega(j_P)-
\tfrac{1}{4}(z^2-z^{-2}) {*\bigl(j_P-\Omega(j_P)\bigr)} \ \\
 &\kern2cm+\tfrac{1}{2}(z^2+z^{-2})j_D-\tfrac{1}{2}(z^2-z^{-2}){*j}_D
\end{aligned}
\end{equation}
and its flatness condition implies the set of equations  \eqref{phase}.
 Given the fact that
after the T-duality (combined with  $Y\mapsto \tilde Y=Y^{-1}$)
we obtain the very same AdS$_5$ sigma model, the
corresponding expression for $ \hj(z)$ in the T-dual model
should be the same as \rf{eq:lax1} with
$(X^{\dot\alpha\beta},Y)\mapsto  (\tilde X^{\dot\alpha\beta},\tilde Y)$.
However,  by  applying the current
duality transformation \eqref{dul} to $\hj(z)$,  we find
\begin{equation}\label{eq:lax2}
 \begin{aligned}
  \tilde \hj(z)\ &=\ \tfrac{1}{4}(z+z^{-1})^2 {*j}_P-\tfrac{1}{4}(z-z^{-1})^2
  {*\Omega(j_P)}-\tfrac{1}{4} (z^2-z^{-2})\left(j_P-\Omega(j_P)\right)\ \\
  &\kern2cm-\tfrac{1}{2}(z^2+z^{-2})j_D+\tfrac{1}{2}(z^2-z^{-2}){*j}_D,
 \end{aligned}
\end{equation}
which does not seem to be the same as  \rf{eq:lax1}
despite the fact that the equations \eqref{phase} are invariant under
\eqref{dul}. Superficially, that may seem to imply that there are two
independent Lax connections with inequivalent monodromy matrices, Yangians, etc.

This, of course, is not the case:   the Lax connections
\eqref{eq:lax1} and \eqref{eq:lax2} are actually  related by a (spectral
parameter dependent)  $\IZ_2$-automorphism of the Lie algebra $\mathfrak{g}$
defined as follows ($T\in\mathfrak{g}$):
\begin{equation}\label{eq:algauto}
 T\ \mapsto\  \CU_z (T) \ :=\ \  U_z \Omega(T)  U^{-1}_z ,
 \qquad\mbox{with}\qquad
 U_z\ :=\  \left(\frac{z-z^{-1}}{z+z^{-1}}\right)^{\im D}.
\end{equation}
This  implies the following action on the components of the current:
\begin{equation}
 \CU_z( j_P)\ =\  \frac{z-z^{-1}}{z+z^{-1}}\ \Omega(j_P),\qquad
 \CU_z( \Omega(j_P)) \ =\ \frac{z+z^{-1}}{z-z^{-1}}\ j_P \qquad\mbox{and}\qquad
 \CU_z( j_D ) \ =\ -j_D,
\end{equation}
and it  is easy to verify that this automorphism maps the two Lax connections into each other
\begin{equation}\label{eq:LaxDuality}
 \CU_z( \hj (z))\ =\ \tilde \hj(z).
\end{equation}
Thus the  T-duality for the bosonic sigma model can be abstractly understood
as a symmetry of the Lax connection (integrable structure)
induced by the automorphism of the conformal algebra $\mathfrak{so}(2,4)$.
This symmetry  then implies a certain map of conserved charges.
We shall make few comments on conserved charges at the end
of Sec.~\ref{sec:CBFD} and in Appendix \ref{sec:CC}.
The present formulation makes the analysis done in \cite{Ricci:2007eq} more transparent.

\

Finding an analogous automorphism once the fermions are included
may not seem   straightforward at first glance. One reason is that a 
particular $\kappa$-symmetry gauge choice makes some of the superisometries
non-manifest. For example, in the T-dual action \eqref{dukt} the original
supersymmetry transformations reduced to  fermionic
shifts of $\theta^{i\alpha}$ and $\bar\theta^{\dot{\alpha}}_i$ (the T-dual
bosonic coordinates $\tilde X^{\dot\alpha\beta}$, being related to
supersymmetric invariants were  not transforming). Furthermore, the above construction
of the automorphism \eqref{eq:algauto}  relied on the fact that
after the  T-duality  we obtain  the very same sigma model action.

In the next section, we will extend the above considerations by combining the bosonic duality
transformation with a certain  fermionic one \cite{BerTalk}. This appears one to require to supplement
the transformation \rf{dul} by a certain transformation (not involving the Hodge star) of the
fermionic components of the current that should produce a symmetry of the full first-order
system \rf{iden}, \rf{eom} written in the $H$-symmetry gauge \rf{eq:rep1}.

It also appears necessary  to consider a different real form of the  complexified
superconformal algebra.
To repeat the above argument about the invariance of the Lax  connection under the duality,
we will then construct an extension of the $\IZ_2$-automorphism \eqref{eq:algauto} to a
$\IZ_4$-automorphism of the full superconformal algebra.

\

\section{Fermionic duality transformation and self-duality of the superstring}\label{sec:last}

The action \rf{dukt} obtained from the gauge-fixed AdS$_5\times S^5$ superstring action \rf{kta}
by the duality transformation applied to the four bosonic coordinates $X$
has  manifest conformal symmetry but not the full superconformal symmetry.
Part of the supersymmetry became non-manifest due to the $\kappa$-symmetry gauge choice\foot{To
recover it one needs to  combine the symmetry transformation with a compensating
$\kappa$-symmetry transformation as in, e.g., the usual light-cone gauge in flat space.}
but part was   made non-local (or trivial) as a result of the duality transformation.
Since the duality is an equivalence transformation at the full $2d$ field theory level,
the original global symmetry  and the associated conserved charges should  not actually
disappear but they may become effectively non-local and thus  hidden (and indeed not visible
in the point-particle limit of the action).

One may ask if one may to  recover the original   global symmetry
in a manifest way, i.e. also at the point-particle level, by combining the bosonic duality
transformation  with a similar one applied to  fermions.
This is   indeed  possible following  the suggestion of \cite{BerTalk}.
As we will  show below, starting with   the  action \rf{dukt}
obtained by the bosonic duality  and applying
a  duality transformation to the    fermionic coordinates  $\theta^{i\alpha}$
(but not to their  conjugates $\bar\theta_{i}^{\dot \alpha}$),
one finds the action that  can be interpreted   as the original
AdS$_5\times S^5$ superstring action written in a different $\kappa$-symmetry gauge.
That means that the combination of the bosonic and the fermionic world-sheet duality transformations
maps not only the bosonic AdS$_5\times S^5$ part, but the  full superstring action into
an equivalent action. As a result, we find the full global superconformal group now acting
(modulo a compensating $\kappa$-symmetry transformation) on coordinates of the dual action.

The fact  that the fermionic duality is performed along the complex (chiral)
fermionic coordinates implies that the resulting action is not Hermitian.
Indeed, to interpret  it as a $\kappa$-symmetry gauge fixed version of the
AdS$_5\times S^5$ superstring action, we will need to formally complexify the
action and choose a special $\kappa$-symmetry (previously considered in \cite{Roiban:2000yy}).

We shall start with a discussion of the superstring action in  this complex
gauge and then show that this action becomes equivalent
to the action \rf{dukt} in the S-gauge upon application of a
fermionic duality transformation. This combined action of bosonic and fermionic dualities
thus maps  the AdS$_5\times S^5$ sigma model into itself.\footnote{This
means, in particular, that this duality transformation induces a map on the space of solutions
of the classical sigma model equations of motion. More precisely, 
we may interpret this T-duality as a dressing transformation acting on the space of solutions,  
like a   B\"acklund transformation (see,  e.g., \cite{Arutyunov:2005nk}).}

We shall then explain the  reason for the fermionic duality transformation by arguing that its
combined action with the bosonic duality is eventually a symmetry of the  first-order system
and of the Lax connection of the superstring model (generalizing a similar symmetry of the 
bosonic model discussed in Sec.~\ref{BB}).

\subsection{Superstring action in a complex \texorpdfstring{$\kappa$}{kappa}-symmetry gauge}\label{sec:RSgauge}

Let us go back to our choice of the coset representative \eqref{eq:rep1}.
Instead of choosing the real S-gauge  \rf{sg}
where $\theta^{i\alpha}_-$  and its conjugate $\bar \theta^{\dot\alpha}_{-i}$
are set to zero,  we may  also consider  the following gauge:
\begin{equation}\label{qs}
 \theta^{i\alpha}_-\ =\ 0\ =\ \bar \theta_{+i}^{\dot\alpha}.
\end{equation}
More precisely, to be able to choose such a
gauge requires a complexification of the AdS$_5\times S^5$
action, i.e.\ a relaxation of  the reality  condition \rf{rea}.
A similar gauge appeared  earlier  in \cite{Roiban:2000yy}
where the authors  considered the superstring action
for a different (Wick rotation related)  slice
of the  complexified version of the AdS$_5\times S^5$ coset superspace \eqref{eq:supercoset}.
A need for such complexification or analytic continuation seems
intimately related to the notion  of dual superconformal
symmetry (cf.~\cite{SokTalk,KorTalk,Drummond:2008vq}).\footnote{From
the field theory point of view,
this complexification  seems to be related to the
PCT self-conjugacy of the $\CN=4$ SYM multiplet which admits a
holomorphic description in the  on-shell superspace \cite{SokTalk,KorTalk,Drummond:2008vq}.} 

In this gauge,  the fermionic part of the coset representative $g= B(X,Y)\eu^{-F(\Theta)}$
in \eqref{eq:rep1} becomes (cf. \rf{krg})
\begin{equation} \la{qsf}
  F(\Theta)\ =\  \im\big[ C_{ij}\epsilon_{\alpha\beta}  \theta^{i\alpha}_+Q^{j\beta}
  +\epsilon_{\dot\alpha\dot\beta}\bar  \theta_{-i}^{\dot\alpha}\bar   S^{i\dot\beta})\big],
\end{equation}
so that  we get a  mixture of $Q$ and $\bar S$ generators
while the $  \bar Q$ and $ S$  parts are  gauged away
(we shall thus refer to this gauge as ${\bar {\rm Q}}$S-gauge).

An interesting feature of this gauge (observed in \cite{Roiban:2000yy})
is that here the  superstring action becomes quadratic
in the fermions even before  T-duality in $X$ as in \cite{Kallosh:1998ji}.\foot{The usual Hermitian
AdS$_5\times S^5$ action in a real $\kappa$-symmetry gauge
can be at best made quartic in the fermions \cite{Kallosh:1998nx,Pesando:1998fv,Metsaev:2000yf,Metsaev:2000yu}.}
Indeed, one may easily verify that here  the current $j$ in \rf{ja} becomes simply
(cf. \eqref{eq:unfixedcurrent})
\begin{equation}
 j\ =\ j_B-\nabla F.
\end{equation}
Going through the  same steps as in Sec.~\ref{sec:KRP}, we then find
\begin{subequations}
\begin{equation}\label{pippo}
 \begin{aligned}
 j\ &=\ \im Y\d
 X^{\dot\alpha\beta}P_{\beta\dot\alpha}+\tfrac{\im}{Y}\d Y D +
   2\im {(\Lambda^{-1})^i}_k\d{\Lambda^k}_j{R_i}^j\ \\
  &\kern1.5cm -\ \im Y^{1/2} {\Lambda^j}_i
        \big(\d\zeta_{j\alpha}+\im
     \d X_{\alpha\dot\beta}\bar  \vartheta^{\dot\beta}_{j}\big)
Q^{i\alpha}
   +\im \epsilon_{\dot\alpha\dot\beta} Y^{-1/2}{\Lambda^j}_i\d\bar  \vartheta^{\dot\beta}_{ j}\bar  
 S^{i\dot\alpha},
\end{aligned}
\end{equation}
where we have defined
\begin{equation}\la{nt}
 \zeta_i^{\alpha}\ :=\ Y^{-1/2} C_{ik}{(\Lambda^{-1})^k}_j\ \theta^{j\alpha}_+
 \qquad\mbox{and}\qquad\bar  \vartheta_{i}^{\dot\alpha}\ :=\ Y^{1/2}{(\Lambda^{-1})^j}_i\
 \bar  \theta^{\dot\alpha}_{-j}.
\end{equation}
\end{subequations}
Extracting the $j_{(m)}$-parts from the above current, the  superstring
action  \rf{eq:action1}  is then found to  take the following explicit form:
\begin{equation}\label{rsa}
\begin{aligned}
 S\ &=\ -\tfrac{T}{2}\int_\Sigma\left[-\tfrac{ 1}{2} W^2
         \d X_{\alpha\dot\beta}\wedge{*\d X}^{\dot\beta\alpha} +
        \tfrac{1}{4 W^2}\,\d  W_{ij}\wedge{*\d} W^{ij}\ \right.\\[-2pt]
    &\kern2.5cm+\ \tfrac{1}{2}   \epsilon^{\alpha\beta} W^{ij}
  (\d\zeta_{i\alpha}+\im\d X_{\alpha\dot\gamma}\bar  \vartheta^{\dot\gamma}_{i})\wedge
 (\d\zeta_{j\beta}+\im\d X_{\beta\dot\delta}\bar  \vartheta^{\dot\delta}_{j})\  \\
     &\left.  \kern3.5cm-\ \tfrac{1}{ 2W^2}\epsilon_{\dot\alpha\dot\beta}
      W^{ij}\d\bar  \vartheta^{\dot\alpha}_{i}\wedge\d\bar  \vartheta_{j}^{\dot\beta}\right]\!.
\end{aligned}
\end{equation}
In deriving this expression, we have used the invariant form
\eqref{eq:cartan-killing} and the
identity $C_{ik}{\Lambda^k}_j=-C_{jk}{\Lambda^k}_i$
and  replaced $Y_{ij}$
by the $SU(4)$  ``rotated'' coordinates $W_{ij}$
(similarly to  $ Y_{ij} \mapsto Z_{ij}$ in \rf{tildeY},  recall that ${\Lambda^i}_j=
Y^{-1} C^{ik} Y_{kj}$)
\begin{equation}
 Y_{ij}\ \mapsto\  W_{ij}\ :=\ Y C_{kl}{(\Lambda^{-1})^k}_i{(\Lambda^{-1})^l}_j
\qquad\mbox{and}\qquad W^2\ =\ \tfrac{1}{4} W_{ij} W^{ij}\ =\ Y^2.  
\end{equation}

\subsection{Fermionic duality transformation}\label{sec:Tfduality}

Let us now  go back to the  T-dual action \eqref{dukt}
found after the bosonic  duality transformation $X \mapsto \tilde X$
in  the superstring action \rf{kta} in the S-gauge
and  show that after the $2d$  duality  applied to  the fermionic
coordinates $\theta^{i\alpha}$ (but not to their  conjugates
$\bar  \theta_i^{\dot \alpha}$)  one finds precisely  the non-Hermitian
 action \rf{rsa}  in the $\bar{\rm Q}$S-gauge  \rf{qs}.

We begin with  the following first-order form of the action \eqref{dukt}:
\begin{equation}
\begin{aligned} \la{nal}
 S\ &=\ -\tfrac{T}{2}\int_\Sigma\Big[-\tfrac{1}{2 Z^2}
         \d\tilde X_{\alpha\dot\beta}\wedge{*\d\tilde X}^{\dot\beta\alpha}
        +\tfrac{1}{4 Z^2}\,\d  Z_{ij}\wedge{*\d} Z^{ij}\ \\
    &\kern2cm-\ \im \tilde X_{\beta\dot\alpha}
      \d\bar \theta^{\dot\alpha}_i\wedge \aA^{i\beta}- \tfrac{1}{2}
      Z_{ij} \aA^{i\alpha}\wedge \aA^j_\alpha
    - \tilde\theta_{i\alpha}\wedge  \d\aA^{i\alpha}
  + \tfrac{1}{2} Z^{ij}\d\bar \theta^{\dot\alpha}_i\wedge\d\bar  \theta_{j\dot\alpha}
\Big]\!,
\end{aligned}
\end{equation}
where we observed that  since \eqref{dukt} depends on $\theta^{i\alpha}$
only through its differential  we can replace $ \d \theta^{i\alpha}$
by $\aA^{i\alpha}$ adding the constraint $\d \aA^{i\alpha}=0$ with
the fermionic Lagrange multiplier $\tilde\theta_{i\alpha}$.
The variation with respect to the gauge potential $\aA^{i\alpha}$ yields
\begin{equation}\la{aa}
 \aA^{i\alpha}\ =\ -\tfrac{1}{ Z^2}
      Z^{ij}\epsilon^{\alpha\beta}(\d\tilde\theta_{j\beta}
  -\im\tilde X_{\beta\dot\alpha}\d\bar \theta^{\dot\alpha}_j) .
\end{equation}
Note that the   on-shell relation
\begin{equation}\la{aab}
 \d \theta^{i\alpha}\ =\ -\tfrac{1}{ Z^2}
      Z^{ij}\epsilon^{\alpha\beta}(\d\tilde\theta_{j\beta}
  -\im\tilde X_{\beta\dot\alpha}\d\bar \theta^{\dot\alpha}_j)
\end{equation}
here is  different compared to the bosonic duality case
\rf{dtr} in that it does not involve the Hodge duality operation. This
has to do with a  peculiarity of the above GS action\foot{For example,
if we compare the two  model Lagrangians $L = f \d \theta \wedge \d \theta $ and
$\tilde L = f^{-1}\d \tilde \theta \wedge \d\tilde \theta $
where $\theta $ and $\tilde \theta$  may carry indices  and $f$  is  a (symmetric)  matrix
depending on worldsheet coordinates then the dual  equations of motion
$  \d f \wedge \d \theta=0$   and  $  \d f^{-1}  \wedge \d\tilde \theta=0$
are first-order  in fermions (as is  standard for the GS string).} where
the fermions  which we dualize appear only in the WZ term.\foot{Let us
mention  that since the fermionic duality can be  performed  via a Gaussian path integral, it
 can be  promoted to a duality of  the quantum sigma model. In particular, when 
performing this duality at the path integral level, one picks up a factor $\Delta_F^{1/2}$
involving the functional determinant $\Delta_F=\Pi_{\sigma\in\Sigma}Z^{16}(\sigma)$. 
Notice that this is the very same functional determinant which already appeared in the bosonic
case (see footnote \ref{foot:shift}). For the combination of bosonic and fermionic dualities
to be promoted to a quantum symmetry of the GS string on AdS$_5\times S^5$, the fermionic 
determinant should then be regularized the same way as the bosonic one, so that 
$\Delta_B^{-1/2}\Delta_F^{1/2}=1$ at the end.}

Substituting  $\aA^{i\alpha}$ in \rf{aa} into \rf{nal}, we end up with the fermionic dual
of this action
\begin{equation}\label{tfd}
\begin{aligned}
 S\ &=\ -\tfrac{T}{2}\int_\Sigma\Big[-\tfrac{1}{2 Z^2}
         \d\tilde X_{\alpha\dot\beta}\wedge{*\d\tilde X}^{\dot\beta\alpha}
        +\tfrac{1}{4 Z^2}\,\d  Z_{ij}\wedge{*\d} Z^{ij}\  \\
    & \kern1.5cm-\
     \tfrac{1}{ 2Z^2} Z^{ij}\epsilon^{\alpha\beta}
       (\d   \tilde\theta'  _{i\alpha}+\im\d\tilde X_{\alpha\dot\gamma}\bar \theta^{\dot\gamma}_i)\wedge
       (\d \tilde\theta'_{j\beta}+\im\d\tilde X_{\beta\dot\delta}\bar \theta^{\dot\delta}_j)
    +  \tfrac{1}{2} Z^{ij}\d\bar \theta^{\dot\alpha}_i\wedge\d\bar \theta_{j\dot\alpha} \Big],
\end{aligned}
\end{equation}
where we have performed the following  fermionic field redefinition:
\begin{equation}
 \tilde\theta'_{i\alpha}
 \ :=\ \tilde\theta_{i\alpha}-\im\tilde X_{\alpha\dot\beta}
     \bar \theta^{\dot\beta}_i.
\end{equation}
Comparing  now the  actions \eqref{rsa} and \eqref{tfd}, we conclude  that
they  coincide provided we make the following  field identifications:
\begin{equation}\la{map}
 X^{\dot\alpha\beta}\ \mapsto\ \tilde X^{\dot\alpha\beta},\quad
 W^{ij}\ \mapsto\ Z^{-2}  Z^{ij},  \quad W\ =\ Y\ \mapsto\  Z^{-1},  \quad
 \zeta_{i\alpha}\ \mapsto\ -\im\tilde\theta'_{i\alpha},\quad
 \bar  \vartheta^{\dot\alpha}_{i}\ \mapsto\ -\im\bar \theta^{\dot\alpha}_i.
\end{equation}
We conclude  that a combination of the bosonic duality \cite{Kallosh:1998ji} and the fermionic duality 
\cite{BerTalk} transformations relates the AdS$_5\times S^5$ superstring action in the supercoset
parametrization \rf{eq:rep1}  and in the $\kappa$-symmetry S-gauge  \rf{sg}
to the same   action in the $\kappa$-symmetry $\bar{\rm Q}$S-gauge  \rf{qs}
(modulo the necessity of  complexification in the transformation process).\foot{Let us
note also that one may consider a more general combinations
of the  bosonic  and fermionic dualities. For example, one may first perform
the fermionic duality and then  the bosonic one; the resulting action will
 be different (and much more complicated). One may also consider combining these dualities  with
linear field redefinitions, getting an analog of the usual $O(d,d)$ duality group.}

This implies  that the original  AdS$_5\times S^5$ action after bosonic and fermionic dualities
has an equivalent (in a  complexified sense) superconformal $PSU(2,2|4)$
global symmetry group, modulo the  fact that some of the supersymmetries are not manifest due
to a special $\kappa$-symmetry gauge choice. In particular, as discussed in \cite{Ricci:2007eq} and above,
(part of \foot{The Lorentz and  the $R$-symmetry $SO(6)$ symmetries  are shared by  the dual
models.}) the corresponding Noether charges of the dual model should have their origin in the
hidden charges  of the original model and vice versa.

In the remainder of this section,  we shall  explain the need
for the fermionic duality transformation from a more general point of view: we will show that  the
combined action of the bosonic and fermionic dualities
leaves the  superstring first order system of equations
and Lax connection invariant generalizing what
 we have done in  the bosonic case   in  Sec.~\ref{BB}

\subsection{Combined bosonic/fermionic duality as a symmetry of the Lax connection}\label{sec:CBFD}

For the bosonic AdS$_5$ sigma model we have shown that the action of T-duality can be interpreted as a
symmetry of first-order system of equations combined  with a particular automorphism of the the
conformal group. In this section we show how to extend that symmetry   to the full superstring
by relating it to  an automorphism of the superconformal algebra.

To start with, we need to extend the action of the
 operator $\Omega$  used in Sec.~\ref{BB} to the full set of the  superconformal generators
\begin{equation}
 \begin{aligned}
  \Omega(P_{\alpha\dot\beta})\ =\ -K_{\alpha\dot\beta},\qquad 
  \Omega(K_{\alpha\dot\beta})\ =\ -P_{\alpha\dot\beta},\qquad
  \Omega(D)\ =\ -D,\qquad 
  \Omega(L_{\alpha\beta})\ =\ L_{\alpha\beta},\\
  \Omega(R_{[ij]})\ = \ -R_{[ij]},\qquad
  \Omega(R_{(ij)})\ = \ R_{(ij)},\\
  \Omega(Q^{i\alpha})\ =\ \im C^{ij}S_j^\alpha,\qquad
  \Omega(\bar Q^{\dot\alpha}_i)\ =\ \im C_{ij}\bar S^{j\dot\alpha},\\
  \Omega(S^\alpha_i)\ =\ -\im C_{ij}Q^{j\alpha},\qquad
  \Omega(\bar S^{i\dot\alpha})\ =\ -\im C^{ij}\bar Q_j^{\dot\alpha}.
 \end{aligned}
\end{equation}
It is easy to verify that $\Omega$ is a $\IZ_4$-automorphism of the
$\mathfrak{psu}(2,2|4)$ algebra.

We can then introduce the following projectors
\begin{equation}
 \begin{aligned}
  \cP_{(0)}\ :=\ \tfrac{1}{4}(1+\Omega+\Omega^2+\Omega^3)\qquad &\mbox{and}\qquad
  \cP_{(2)}\ :=\ \tfrac{1}{4}(1-\Omega+\Omega^2-\Omega^3),\\
  \cP_{(1)}\ :=\ \tfrac{1}{4}(1-\im\Omega-\Omega^2+\im\Omega^3)\qquad &\mbox{and}\qquad
  \cP_{(3)}\ :=\ \tfrac{1}{4}(1+\im\Omega-\Omega^2-\im\Omega^3),
 \end{aligned}
\end{equation}
which give the $\IZ_4$-decomposition of the algebra
\begin{equation}
 \mathfrak{g}_{(m)}\ =\ \cP_{(m)}(\mathfrak{g})
\end{equation}
presented in \eqref{eq:explicitsplitting}. From the above analysis we know that the
superstring  action in the  S-gauge is mapped under the combined action of the  bosonic
and fermionic dualities into the superstring  action in the $\bar{\rm Q}$S complex gauge.
To understand this relation from more general perspective, let us start by presenting the
$\IZ_4$-decomposition of the currents in the S-gauge
\begin{equation}
 \begin{aligned}
  j_{(0)}\ =\ A\ =\ \tfrac{1}{2}(1+\Omega)(j_P+j_R)&\qquad\mbox{and}\qquad
  j_{(2)}\ =\ \tfrac{1}{2}(1-\Omega)(j_P+j_R+j_D),\\
  j_{(1)}\ =\ \tfrac{1}{2}(1-\im\Omega)(j_Q+j_{\bar Q})&\qquad\mbox{and}\qquad
  j_{(3)}\ =\ \tfrac{1}{2}(1+\im\Omega)(j_Q+j_{\bar Q}).
 \end{aligned}
\end{equation}
As follows from \eqref{dtr} and \eqref{aab}, the combined
duality is equivalent to the following action on the superstring fields:
\begin{equation}\label{span}
 \begin{aligned}
  \d X^{\dot\beta\alpha}+\tfrac{\im}{2}(\bar\theta^{\dot\beta}_i\d\theta^{i\alpha}-
  \d\bar\theta^{\dot\beta}_i\theta^{i\alpha})\ &=\   Z^{-2}\,{*\d}\tilde X^{\dot\beta\alpha},\\
  \d \theta^{i\alpha}\ &=\ -\tfrac{1}{Z^2}Z^{ij}\epsilon^{\alpha\beta}(\d\tilde\theta_{j\beta}
  -\im\tilde X_{\beta\dot\alpha}\d\bar \theta^{\dot\alpha}_j).
 \end{aligned}
\end{equation}
In order to be able to compare currents before and after the duality, let us  also change
coordinates according to
\begin{equation}\label{eq:lastcoc}
 \begin{aligned}
  Z_{ij}\ &\mapsto\ \tilde Y^{-2} C_{ik}C_{jl}\tilde Y^{kl},\qquad\mbox{with}\qquad
  \tilde Y^2\ =\ \tfrac{1}{4}\tilde Y_{ij}\tilde Y^{ij}\ =\ Z^{-2},\\
  \tilde\theta_{i\alpha}\ &\mapsto \ -\im(\zeta_{i\alpha}+\im\tilde X_{\alpha\dot\beta}
  \bar \vartheta^{\dot\beta}_i)\qquad\mbox{and}\qquad
  \bar\theta^{\dot\alpha}_i\ \mapsto\ -\im\bar\vartheta^{\dot\alpha}_i.
\end{aligned}
\end{equation}
Upon applying the duality and the coordinate transformation as
above, we can relate the components of the current in
the S-gauge, $j=j_P+j_D+j_R+j_Q+j_{\bar Q}$, to the dual one in the $\bar{\rm  Q}$S-gauge,
$\tilde j=\tilde j_P+\tilde j_D+\tilde j_R+\tilde j_Q+\tilde j_{\bar S}$ as
 follows (see also \eqref{eq:fullcurrent} and \eqref{pippo}):
\begin{equation}
 \begin{aligned}
  j_P\ &=\ \im Y\Pi^{\dot\alpha\beta} P_{\beta\dot\alpha}\ =\
           \tilde Y {*\d}\tilde X^{\dot\alpha\beta} P_{\beta\dot\alpha}\ =\ {*\tilde j_P},\\
  j_D\ &=\ \tfrac{\im}{Y}\d YD\ =\ -\tfrac{\im}{\tilde Y} \d \tilde Y D\ =\ -\tilde j_D,\\
  j_{R_a}\ &=\  -  2\im C^{l[i}{(\Lambda^{-1}(Y))^{j]}}_k\d{\Lambda(Y)^k}_lR_{[ij]}
              \ =\  2\im C^{l[i}{(\Lambda^{-1}(\tilde Y))^{j]}}_k\d{\Lambda(\tilde Y)^k}_lR_{[ij]}
              \ =\ -\tilde j_{R_a},\\
  j_{R_s}\ &=\  -  2\im C^{l(i}{(\Lambda^{-1}(Y))^{j)}}_k\d{\Lambda(Y)^k}_lR_{(ij)}
              \ =\ -  2\im C^{l(i}{(\Lambda^{-1}(\tilde Y))^{j)}}_k\d{\Lambda(\tilde Y)^k}_lR_{(ij)}
              \ =\ \tilde j_{R_s},\\
  j_Q\ &=\  -\im C_{ij}\epsilon_{\alpha\beta}Y^{1/2} {\Lambda(Y)^i}_k\d\theta^{k\alpha}Q^{j\beta}
              \ =\ - \tilde Y^{1/2} {\Lambda(\tilde Y)^j}_i\big(\d\zeta_{j\alpha}+\im
                 \d\tilde X_{\alpha\dot\beta}\bar  \vartheta^{\dot\beta}_{j}\big)
                 Q^{i\alpha}\ =\ -\im\tilde j_Q,\\
  j_{\bar Q}\ &=\ \im C^{ij}\epsilon_{\dot\alpha\dot\beta}
                Y^{1/2} {(\Lambda^{-1}(Y))^k}_i\,\d\bar\theta^{\dot\alpha}_{k}\bar Q^{\dot\beta}_j
             \ =\  C^{ij}\epsilon_{\dot\alpha\dot\beta} \tilde Y^{-1/2}{\Lambda(\tilde Y)^k}_i\d\bar  
               \vartheta^{\dot\alpha}_{k}\bar Q_j^{\dot\beta}\ =\ -\Omega(\tilde j_{\bar S}).
 \end{aligned}
\end{equation}
Here, $R_a$ represents $R_{[ij]}$ while $R_s$ represents $R_{(ij)}$.
We can therefore formally summarize the action of
the combined bosonic and fermionic dualities (including the coordinate
transformation \eqref{eq:lastcoc}) on the current as
\begin{equation}\label{fulldual}
 \begin{aligned}
  &j_P\ \mapsto\ \tilde j_P\ =\ {* j_P},\quad && j_D\ \mapsto\ \tilde j_D\ =\ - j_D,\\
  &j_{R_a}\ \mapsto\ \tilde j_{R_a}\ =\ - j_{R_a},\quad && j_{R_s}\ \mapsto\ \tilde j_{R_s}\ =\ j_{R_s},\\
  &j_Q\ \mapsto \tilde j_Q\ =\ \im j_Q,\quad && j_{\bar Q}\ \mapsto\ \tilde j_{\bar S}\ =\ \Omega(j_{\bar Q}).
 \end{aligned}
\end{equation}
The family of flat currents or Lax connection in the S-gauge is
\begin{equation}\label{slax1}
 \hj(z)\ =\ \hj_B(z)+\tfrac{1}{2}(z+z^{-1})(j_Q+j_{\bar Q})
 -\tfrac{\im}{2}(z-z^{-1})(\Omega(j_Q)+\Omega(j_{\bar Q})),
\end{equation}
where $\hj_B(z)$  is formally the current in \eqref{eq:lax1}, with $j_P$ given in \eqref{eq:fullcurrent}. 
Upon applying the duality transformation \eqref{fulldual}, we
 obtain the dual flat current family
\begin{equation}\label{slax2}
 \tilde\hj(z)\ =\ \tilde\hj_B(z)+\tfrac{\im}{2}(z+z^{-1})(j_Q-\im 
 \Omega(j_{\bar Q}))+\tfrac{1}{2}(z-z^{-1})(\Omega(j_Q)+\im j_{\bar Q}),
\end{equation}
where $\tilde\hj_B(z)$ is the current in \eqref{eq:lax2}.

As in the bosonic case, we  obtain two
seemingly different  Lax connections. However, one can  show
that the  two Lax connections \eqref{slax1} and \eqref{slax2}
are again related by a spectral parameter dependent automorphism
of the superconformal algebra. Indeed, we can define  the following
$\IZ_4$-automorphism:
\begin{equation}\label{eq:superauto2}
 T\ \mapsto\ \CU_z(T)\ :=\  U_z\,\Omega(T)\,
 U_z^{-1}\, ,
\qquad\mbox{with}\qquad U_z\ :=\
 \left(\frac{z-z^{-1}}{z+z^{-1}}
\right)^{\im (\BB+D)},
\end{equation}
where $T$ is a generic generator of the
 superconformal algebra and $\BB$ generates a
$U(1)$-automor\-phism, with  non-vanishing (anti-)commutators
being
\begin{equation}
 [\BB,Q]\ =\ \tfrac{\im}{2} Q,\qquad [\BB,S]\ =\
-\tfrac{\im}{2}S,\qquad
[\BB,\bar Q]\ =\ -\tfrac{\im}{2} \bar Q, \qquad
[\BB,\bar S]\ =\ \tfrac{\im}{2} \bar S
\end{equation}
and $\Omega(\BB)=-\BB$.

If we define
\begin{equation}
 f(z)\ :=\ \frac{z-z^{-1}}{z+z^{-1}},
\end{equation} 
we can represent the explicit action of the automorphism $\CU_z$ on the generators as follows
\begin{equation}
 \begin{aligned}
  & \CU_z(P_{\alpha\dot\beta})\ =\ f(z)\Omega(P_{\alpha\dot\beta}), \qquad \CU_z(K_{\alpha\dot\beta})
     \ =\ f^{-1}(z)\Omega(K_{\alpha\dot\beta}),\qquad \CU_z(D)\ =\ \Omega(D),\\
  & \CU_z(R_i^j)\ =\ \Omega(R^j_i), \ \qquad \CU_z(L_{\alpha\beta})\ =\ \Omega(L_{\alpha\beta}),\qquad
     \CU_z(L_{\dot\alpha\dot\beta})\ =\ \Omega(L_{\dot\alpha\dot\beta}),\\
  & \CU_z(Q^{i\alpha})\ =\ f(z)\Omega(Q^{i\alpha}),\ \  \qquad \CU_z(S^{\alpha}_i)\ =\ 
      f(z)^{-1} \Omega(S^{\alpha}_i),\\
  & \CU_z(\bar Q^{\dot \alpha}_i)\ =\ \Omega(\bar Q^{\dot \alpha}_i),\qquad 
     \CU_z(\bar S^{i\dot \alpha})\ =\  \Omega(\bar S^{i\dot \alpha}).
 \end{aligned}
\end{equation}
The action of this automorphism on the bosonic generators $\{P,L,K,D\}$ of the $\mathfrak{so}(2,4)$ algebra reduces to the action of the $\IZ_2$-automorphism considered before in Sec.~\ref{BB}
Altogether, we end up with
\begin{equation}
 \tilde \hj(z)\ =\ \CU_z(\hj(z)).
\end{equation}

The conclusion is that  we may interpret the combined action of the
bosonic and the fermionic dualities as a symmetry of the first-order superstring
system of equations induced by the above automorphism of
the (complexified) $\mathfrak{psu}(2,2|4)$ algebra.

Furthermore,  since the Noether charges may be derived from the flat current
$\hj(z)$ near $z=\pm 1$ (see Eqs.~\eqref{eq:ginvflatcur}, \eqref{charge} and also
Appendix \ref{sec:CC} for more details), Eq.~\eqref{eq:superauto2} suggests that the Noether
charges associated with the generators of the superconformal algebra behave under the combined action
of the bosonic and fermionic dualities as follows (modulo the issue of boundary conditions):
\begin{itemize}\setlength{\itemsep}{-3pt}
 \item \ the $P_{\alpha\dot\beta}$-charge becomes trivial
 \item \ the $L_{\alpha\beta}$- and $L_{\dot\alpha\dot\beta}$-charges go 
         into themselves and thus remain local
 \item \ the $K_{\alpha\dot\beta}$-charge gets lifted and becomes non-local
 \item \ the $D$-charge goes into itself and thus remains local
 \item \ the ${R_i}^j$-charge goes into itself and thus remains local
 \item \ the $Q^{i\alpha}$-charge becomes trivial
 \item \ the $\bar Q^{\dot\alpha}_i$-charge goes into the $\bar S^{i\dot\alpha}$-charge and thus
         remains local
 \item \ the  $S_i^{\alpha}$-charge gets lifted and becomes non-local
 \item \ the  $\bar S^{i\dot\alpha}$-charge goes into the $\bar Q^{\dot\alpha}_i$-charge and thus
         remains local
\end{itemize}
\centerline{\small {\bf Table 5.1:} Behavior of the Noether charges under bosonic and fermionic dualities.}
\vspace*{8pt}
This is an immediate consequence of the fact that $f$ goes to zero near $z=\pm1$
while $f^{-1}$ diverges as can be seen from the respective expansions around $z=\pm1$
\begin{equation}
\begin{aligned}
 f(z)\ &=\ \pm(z\mp1)+\CO((z\mp1)^2)\ \sim\ 0\qquad\mbox{for}\qquad z\ \to\ \pm1,\\
 f^{-1}(z)\ &=\ \pm\tfrac{1}{z\mp1}+\tfrac{1}{2}+\CO((z\mp1))\
\sim\ \pm\tfrac{1}{z\mp 1}\qquad\mbox{for}\qquad z\  \to\ \pm1.
\end{aligned}
\end{equation}
We can understand the behavior of $P_{\alpha\dot\beta}$
and $Q^{i\alpha}$ under T-duality also by observing that they do not act on the
dual coordinates $\tilde X_{\alpha\dot\beta}$ and $\tilde\theta_{i\alpha}$ given in
Eq.~\eqref{span}. This is what we mean by `trivial' in the above list.
The resulting picture   is  in agreement with the
conclusion announced  in \cite{BerTalk}. Remarkably, similar relations
for  the generators of the original and dual
superconformal symmetry when acting on supergluon amplitudes
appear also on the gauge theory side \cite{SokTalk,KorTalk,Drummond:2008vq}.\foot{Some generators act
trivially and some do not act linearly
as they are realized as second-order differential operators \cite{SokTalk,KorTalk,Drummond:2008vq}. In this reference, the 
amplitudes are discussed in a chiral superspace. An equivalent choice would have been  to consider
an anti-chiral superspace. In the present discussion this change would amount
 to choose  a Q$\bar{\rm S}$-gauge 
rather than a $\bar{\rm Q}$S-gauge and to perform the  fermionic 
duality along the $\bar\theta$- instead of the $\theta$-directions.}

Let us add  that the automorphism \eqref{eq:superauto2} can, in principle,  be used to obtain a 
map between the full set of conserved charges (local and non-local ones) before and after the duality.

\

It would be useful to give a more covariant version of
the above  analysis in which the $\kappa$-symmetry would not be fixed. This would make
the global symmetries more manifest and would further clarify the  mapping between the
conserved charges in the two dual models. It would also be interesting to understand further
the meaning of complexification of the superconformal algebra which was required in our string
theory considerations and which  apparently is also playing an  important role on the dual gauge
theory side \cite{SokTalk,KorTalk,Drummond:2008vq} (being related  to a possibility of having  a chiral on-shell superspace
description of the scattering amplitudes for the PCT self-conjugate  $\CN=4$ SYM multiplet).

\

Needless to say, the major outstanding problem is to understand the precise relation between
the superstring symmetries in the bulk and the symmetries of the
supergluon scattering amplitudes in the boundary gauge theory.
This would presumably require defining the IR-regularized amplitudes in terms of
correlators of open-string vertex operators inserted on an IR D3-brane as in \cite{Alday:2007hr,Alday:2007he}
(see also \cite{Alday:2008yw} for a review). For that, one would have to specify, in particular,
 the boundary conditions  for the open strings stretching in the bulk of AdS$_5$ and ending on
the IR brane. The presence of the IR regulator would break the (dual) superconformal symmetry,
but in an anomalous, i.e.\ ``controlled'',  way \cite{SokTalk,KorTalk,Drummond:2008vq}: it will still lead to highly non-trivial
constraints on the finite parts of the amplitudes.\footnote{In \cite{SokTalk,KorTalk,Drummond:2008vq},
 the superconformal algebra is extended by a central charge 
which is suggested to be related to the helicity of the particles participating
 in the scattering process. To recover this central extension in the present approach, 
 one has, presumably, to consider the action of the superconformal algebra on the (super)gluon
vertex operators  which define the scattering amplitude  on the string theory side.}

\

\vspace*{.5cm}
\noindent
{\bf Note  added.}
While this paper was being prepared for submission, we received a draft of the  forthcoming
paper \cite{Berkovits:2008ic} (some of the results of which were announced  in \cite{BerTalk}) which  has an overlap 
with the present paper.

\

\vspace*{.5cm}
\noindent
{\bf Acknowledgements.}
We are grateful to F.~Alday, G.~Arutyunov, N.~Berkovits, J.~Drummond, J.~Henn, G.~Korchemsky, 
J.~Maldacena,  R.~Roiban and E.~Sokatchev for important discussions, questions   and suggestions.
M.W.~was supported in part by the STFC under the rolling grant PP/D0744X/1.

\

\appendix

\section{Spinor conventions}\label{sec:gammamatrix}

\subsection*{Four-dimensional spinor conventions}
We mostly follow the conventions of Wess and Bagger \cite{Wess:1992cp}.
Consider 4-dimensional Minkowski space $\IR^{1,3}$
with metric $(\eta_{ab})=\mbox{diag}(-1,1,1,1)$
and coordinates $X^a$, where $a,b,\ldots=0,\ldots,3$.
We shall adopt  the convention  $(\psi_\alpha )^\dagger
= \bar \psi_{\dot \alpha}$ (we shall use `$^\dagger$' to denote Hermitian conjugation on
Gra{\ss}mann algebra elements),
where $\alpha,\beta,\ldots=1,2$ and
$\dot\alpha,\dot\beta,\ldots=\dot1,\dot2$.

Let
$(\sigma^a):=(\sigma^a_{\alpha\dot\beta}):=(-\mathbbm{1}_2,\vec\sigma)$. Here,
$\vec\sigma =(\sigma^1,\sigma^2,\sigma^3)$ are the Pauli matrices.
Then we define $\bar\sigma^{a\dot\alpha\beta}:=\epsilon^{\dot\alpha\dot\gamma}\epsilon^{\beta\delta}
\sigma^a_{\delta\dot\gamma}$, with $\epsilon_{\alpha\gamma}\epsilon^{\gamma\beta}={\delta_{\alpha}}^\beta$,
$\epsilon_{\dot\alpha\dot\gamma}\epsilon^{\dot\gamma\dot\beta}={\delta_{\dot\alpha}}^{\dot\beta}$
and $\epsilon_{12}=-\epsilon_{21}=\epsilon_{\dot1\dot2}=-\epsilon_{\dot2\dot1}=1$.
Next we introduce
\begin{equation}
 X_{\alpha\dot\beta}\ :=\ \sigma^a_{\alpha\dot\beta} X_a\qquad\Longleftrightarrow\qquad
 X^a\ =\ -\tfrac{1}{2}\bar\sigma^{a\dot\alpha\beta}X_{\beta\dot\alpha},
\end{equation}
where $X^a=\eta^{ab}X_b$.
Explicitly, this reads as
\begin{equation}
 (X^{\dot\alpha\beta})\ =\ \begin{pmatrix}
                             X^0-X^3 & -X^1+\im X^2\\ -X^1-\im X^2 & X^0+X^3
                           \end{pmatrix}\!.
\end{equation}
 The Minkowski space  line element is then given by
\begin{equation}
 \d s^2\ =\ -\det(\d X^{\dot\alpha\beta})\ =\ -\tfrac{1}{2}\epsilon_{\alpha\beta}\epsilon_{\dot\gamma\dot\delta}
    \d X^{\dot\gamma\alpha}\d X^{\dot\delta\beta}\ =\ -\tfrac{1}{2}\d X_{\alpha\dot\beta}\d X^{\dot\beta\alpha}.
\end{equation}
From the Hermiticity of $\sigma^a$ and $\bar\sigma^a$, i.e.\ $\sigma^a=(\sigma^a)^\dagger$ and
$\bar\sigma^a=(\bar\sigma^a)^\dagger$ it follows that
\begin{equation}
 X_{\alpha\dot\beta}\ =\ (X_{\beta\dot\alpha})^*\qquad\mbox{and}\qquad
 X^{\dot\alpha\beta}\ =\ (X^{\beta\dot\alpha})^*.
\end{equation}

\subsection*{Six-dimensional spinor conventions}
Consider 6-dimensional Euclidean space $\IR^6$ with metric
$\delta_{rs}$ and coordinates $Y^r$, where $r,s,\ldots=1,\ldots,6$. Then
we take the  $\sigma^r$ and
$\bar\sigma^r$ matrices  as anti-symmetric $4\times4$
matrices,
\begin{equation}\label{eq:6dmetric}
\sigma^r\ =\ (\sigma^r_{ij}),\qquad\sigma^r_{ij}\ =\ -\sigma^r_{ji}\qquad\mbox{and}\qquad
 \bar\sigma^{rij}\ :=\ \tfrac{1}{2}\epsilon^{ijkl}\sigma^r_{kl},
\end{equation}
for $i,j,\ldots=1,\ldots,4$ and
$\epsilon^{ijkl}$ is totally anti-symmetric in its indices with $\epsilon^{1234}=1$.

Next we introduce
\begin{equation}
 Y_{ij}\ :=\ \sigma^r_{ij} Y_r\qquad\Longleftrightarrow\qquad Y^r\ =\ \tfrac{1}{4}\bar\sigma^{rij}Y_{ij}.
\end{equation}
Furthermore, by virtue of \eqref{eq:6dmetric}  and  $\bar\sigma^{rij}=(\sigma^r_{ij})^*$   we have
\begin{equation}
Y^{ij}\ =\ \tfrac{1}{2}\epsilon^{ijkl}Y_{kl}\qquad\mbox{and}\qquad Y^{ij}\ =\ (Y_{ij})^*.
\end{equation}
The line element of $\IR^6$   in these coordinates is given by
\begin{equation}
 \d s^2\ =\ \sqrt{\det(\d Y^{ij})}\ =\
  \tfrac{1}{8}\epsilon_{ijkl}\d Y^{ij}\d Y^{kl}\ =\ \tfrac{1}{4}\d Y_{ij}\d Y^{ij}.
\end{equation}

\

\section{Superconformal algebra}\label{sec:SCA}

\subsection*{Commutation relations}
The non-trivial (anti-)commutation relations
among the generators
of the superconformal algebra $\mathfrak{psu}(2,2|4)$ are
\begin{equation}
\begin{aligned}
 \{Q^{i\alpha},\bar Q^{\dot\beta}_j\}\ =\ -\delta^i_j P^{\dot\beta\alpha},\qquad
 \{S^{\alpha}_i,\bar  S^{j\dot\beta}\}\ =\ -\delta_i^j K^{\dot\beta\alpha},\\
 \{Q^{i\alpha},S^\beta_j\}\ =\ -\im\delta^i_j(L^{\alpha\beta}+\tfrac{1}{2}\epsilon^{\alpha\beta}D)
   +2\im\epsilon^{\alpha\beta}{R_j}^i,\\
 {[{R_i}^j,S^\alpha_k]}\ =\ -\tfrac{\im}{2}(\delta^j_k S^{\alpha}_i-\tfrac{1}{4}\delta^i_jS^\alpha_k),\\
 {[L^{\alpha\beta},S^\gamma_i]}\ =\ \im\epsilon^{\gamma(\alpha}S^{\beta)}_i,\qquad
 {[P^{\dot\alpha\beta},S^\gamma_i]}\ =\ \epsilon^{\beta\gamma}\bar Q^{\dot\alpha}_i,\qquad
  {[D,S^\alpha_i]}\ =\ -\tfrac{\im}{2}S^\alpha_i,\\
 {[{R_i}^j,Q^{k\alpha}]}\ =\ \tfrac{\im}{2}(\delta^k_iQ^{j\alpha}-\tfrac{1}{4}\delta^j_iQ^{k\alpha}),\\
 {[L^{\alpha\beta},Q^{i\gamma}]}\ =\ \im\epsilon^{\gamma(\alpha}Q^{i\beta)},\qquad
 {[K^{\dot\alpha\beta},Q^{i\gamma}]}\ =\ \epsilon^{\beta\gamma}\bar  S^{i\dot\alpha},\qquad
 {[D,Q^{i\alpha}]}\ =\ \tfrac{\im}{2}Q^{i\alpha},\\
 {[{R_i}^j,{R_k}^l]}\ =\ \tfrac{\im}{2}(\delta^l_i{R_k}^j-\delta^j_k{R_i}^l),\\
 {[D,P^{\dot\alpha\beta}]}\ =\ \im P^{\dot\alpha\beta},\qquad
 {[D,K^{\dot\alpha\beta}]}\ =\ -\im K^{\dot\alpha\beta},\\
 {[L^{\alpha\beta},P^{\dot\gamma\delta}]}\ =\ \im\epsilon^{\delta(\alpha}P^{\dot\gamma\beta)},\qquad
 {[L^{\alpha\beta},K^{\dot\gamma\delta}]}\ =\ \im\epsilon^{\delta(\alpha}K^{\dot\gamma\beta)},\\
 {[L_{\alpha\beta},L^{\gamma\delta}]}\ =\ -2\im\delta_{(\alpha}^{(\gamma}L_{\beta)}^{\delta)},\\
 {[P^{\dot\alpha\beta},K^{\dot\gamma\delta}]}\ =\ -\im(\epsilon^{\dot\alpha\dot\gamma}L^{\beta\delta}
  +\epsilon^{\beta\delta}L^{\dot\alpha\dot\gamma}+\epsilon^{\beta\delta}\epsilon^{\dot\alpha\dot\gamma}D).
\end{aligned}
\end{equation}
In writing these expressions, we have made use of the $4d$ vector
index identification $\{a\}=\{\alpha\dot\beta\}$.
In particular, this
implies that the rotation generators decompose into the self-dual and anti-self-dual parts
\begin{equation}
 L_{ab} \ \leftrightarrow \ L_{\alpha\dot\beta\gamma\dot\delta}\ =\ 
 -\tfrac{1}{2}(\epsilon_{\dot\beta\dot\delta}L_{\alpha\gamma}+
    \epsilon_{\alpha\gamma}L_{\dot\beta\dot\delta}), \
    \ \ \ \ \ L_{\alpha\beta}\ =\ L_{\beta\alpha} , \ \ \ \ \ \
    L_{\dot\alpha\dot\beta}\ =\ (L_{\alpha\beta})^\dagger  .
\end{equation}

\subsection*{Invariant form on $\mathfrak{psu}(2,2|4)$}

The non-vanishing components of the invariant form of
$\mathfrak{psu}(2,2|4)$ compatible with the above choice of the basis  of the algebra are
\begin{equation}\label{eq:cartan-killing}
\begin{aligned}
  \mbox{str}(P_{\alpha\dot\beta}K_{\gamma\dot\delta})\ =\ \epsilon_{\alpha\gamma}\epsilon_{\dot\beta\dot\delta},
  \qquad\mbox{str}(DD)\ =\ -1,\qquad\mbox{str}(L_{\alpha\beta}L_{\gamma\delta})\ =\ -\epsilon_{\alpha(\gamma}
  \epsilon_{\delta)\beta},\\
  \mbox{str}({R_i}^j{R_k}^l)\ =\ \tfrac{1}{4}(\delta^l_i\delta^j_k-\tfrac{1}{4}\delta^i_j\delta^l_k),\qquad
  \mbox{str}(Q^{i\alpha}S^\beta_j)\ =\ \delta^i_j\epsilon^{\alpha\beta}.\kern1.5cm
\end{aligned}
\end{equation}

\subsection*{$\IZ_4$-grading of the algebra}

With the  above choice  of the generators,
the $\IZ_4$-decomposition \eqref{eq:algsplit} is not manifest. To find
a manifest realization of the grading, let us start from the bosonic
part of the algebra, in particular from $\mathfrak{so}(6)\cong\mathfrak{su}(4)=\mbox{span}\{{R_i}^j\}$.
Since $S^5\cong SO(6)/SO(5)\cong SU(4)/Sp(4)$, we may pick some
$Sp(4)$-metric $C_{ij}$ with
\begin{equation}
 C_{ij}\ =\ -C_{ji}\ =:\ \tfrac{1}{2}\epsilon_{ijkl}C^{kl}, \qquad
 C_{ij}\ =\ (C^{ij})^*\, \qquad\mbox{and}\qquad  C_{ik}C^{jk}\ =\ \delta^j_i.
\end{equation}
Without loss of generality,
$C=(C_{ij})$ may be chosen  as
\begin{equation}
 C\ =\ \begin{pmatrix}
        0 & 1 & 0 & 0\\ -1 & 0 & 0 & 0\\ 0 & 0 & 0 & -1\\ 0 & 0 & 1 & 0
       \end{pmatrix}\!.
\end{equation}
A particular  choice of $C_{ij}$ induces an isomorphism (non-canonical)
between the 15-dimensional bi-vector
representation ${\bf 6}\wedge{\bf 6}$ of $\mathfrak{so}(6)\cong\mathfrak{su}(4)$
and the sum of the 5-dimensional vector representation ${\bf 5}$ of
$\mathfrak{so}(5)\cong\mathfrak{sp}(4)\subset\mathfrak{so}(6)$ and
the 10-dimensional bi-vector representation ${\bf 5}\wedge{\bf 5}$, i.e.
${\bf 6}\wedge{\bf 6}\cong{\bf 5}\oplus{\bf 5}\wedge{\bf 5}$. Explicitly,
we then have
\begin{equation}
 \underbrace{R_{ij}\ :=\ C_{ik}{R_j}^k}_{\hat=\, {\bf 6}\wedge{\bf 6}}
  \ =\ \underbrace{C_{[ik}{R_{j]}}^k}_{\hat=\, {\bf 5}}+
 \underbrace{C_{(ik}{R_{j)}}^k}_{\hat=\, {\bf 5}\wedge{\bf 5}}.
\end{equation}
Note that $C_{[ik}{R_{j]}}^k$ represents the ${\bf 5}$  because of ${R_i}^i=0$.
We shall use the  notation
 \begin{equation} 
  R_{(ij)}\ :=\ C_{(ik}{R_{j)}}^k\qquad\mbox{and}\qquad  
  R_{[ij]}\ :=\ C_{[ik}{R_{j]}}^k,
\end{equation}
where  parentheses (square brackets) mean normalized (anti-)symmetrization.
Then $R_{(ij)}\in\mathfrak{h}$
and $R_{[ij]}\in\mathfrak{g}_{(2)}$.

Next,  consider $\mathfrak{so}(2,4)\cong\mathfrak{su}(2,2)$ which is generated by $P_a,L_{ab},K_a,D$.
Then $\frac12(P_a-K_a)$ and $L_{ab}$ are the remaining generators of
$\mathfrak{h}$ while $\frac12(P_a+K_a)$ and $D$ are the remaining generators
of the bosonic coset part  $\mathfrak{g}_{(2)}$, respectively.
One may proceed similarly with the
fermionic generators. Eventually, one finds that the $\IZ_4$-splitting is given
by \eqref{eq:explicitsplitting}.

\

\section{Fermionic current}\label{sec:CR}

Here,  we shall briefly review the derivation of Eq.~\eqref{eq:unfixedcurrent}. The bosonic current $j_B$ was already given.
To get a handle on the fermionic one $j_F$, let us consider the one-parameter family ($t\in\IR$)
\begin{equation}
 j(t)\ :=\ \eu^{tF} j_B \,\eu^{-tF}+\eu^{tF}\d \eu^{-Ft},\qquad\mbox{with}\qquad
 j(t=0)\ =\ j_B .
\end{equation}
This then implies
\begin{equation}
 \partial_t j(t)\ =\ \eu^{Ft} (-\nabla F)\,\eu^{-tF},\qquad\mbox{with}\qquad
 \nabla\,\cdot\ =\ \d\cdot+[j_B ,\,\cdot\,]
\end{equation}
and so $(j(t=1)=j$)
\begin{equation}
 j(t)\ =\ j_B + \int_0^t\d t'\, \eu^{t'F} (-\nabla F)\,\eu^{-t'F}\qquad\Longrightarrow\qquad
 j_F\ =\ \int_0^1\d t'\, \eu^{ t'F} (-\nabla F)\,\eu^{-t'F}.
\end{equation}
Upon recalling the formula $\eu^{tA} B \eu^{-tA}=\sum_{n}\frac{t^n}{n!}[A,B]^{(n)}$, where
$[A,B]^{(n)}:=[A,[A,B]^{(n-1)}]$ with $[A,B]^{(0)}:=B$, we find
\begin{equation}\label{eq:fercur}
\begin{aligned}
 j_F \ &=\ \sum_n\tfrac{1}{(n+1)!}[F,-\nabla F]^{(n)}\\
       &=\
        \sum_n\tfrac{1}{(2n+1)!}[F,-\nabla F]^{(2n)}+
        \sum_n\tfrac{1}{(2n+2)!}[F,-\nabla F]^{(2n+1)}.
\end{aligned}
\end{equation}
Using the definition \eqref{eq:defofM}, then a short calculation reveals that
\begin{subequations}
\begin{eqnarray}
 \sum_n\tfrac{1}{(2n+1)!}[F,-\nabla F]^{(2n)}
  &=& -\tfrac{\sinh(\mathcal{M})}{\mathcal{\CM}}\,\nabla F,\\
 \sum_n\tfrac{1}{(2n+2)!}[F,-\nabla F]^{(2n+1)}
  &=& -2\left[\im f, \tfrac{\sinh^2(\mathcal{M}/2)}{\mathcal{\CM}^2}\,\nabla F\right]\!.
\end{eqnarray}
\end{subequations}
Altogether, we then obtain \eqref{eq:unfixedcurrent}.

\

\section{Comments on conserved charges}\label{sec:CC}

Let us make some comments  on the construction of
conserved charges for the bosonic sigma model discussed in Sec.~\ref{BB}
While having in mind the  relation to scattering amplitudes in the
dual SYM theory it would be natural to discuss the T-duality acting 
on the open strings as in \cite{Alday:2007hr,Alday:2007he} here we shall formally
assume that the string coordinates are periodic
in the spatial worldsheet direction $\sigma$ as would be the case
in the closed string  sector of the theory.

We have  seen in Sec.~\ref{sec:1pFC} that conserved charges  follow 
directly  from the current $\hJ(z)$. An alternative route to find them
is to consider the parallel transport of the Lax connection $\hj(z)$
\begin{equation}
M(z;\sigma,\tau;\sigma_0,\tau_0)\ :=\ P \exp\left(\int_{\sigma_0,\tau_0}^{\sigma,\tau} \hj(z)\right)\!.
\end{equation}
A candidate conserved charge is given by the following composition of parallel transports
\begin{equation}
Q(z)\ :=\
M(z;\sigma_0,\tau_0;\sigma+2\pi,\tau)\,
\frac{\partial M}{\partial z}(z;\sigma+2\pi,\tau;\sigma,\tau)\,
M(z;\sigma,\tau;\sigma_0,\tau_0).
\end{equation}
Assuming that the Lax connection is periodic, $\hj(z;\sigma+2\pi,\tau)=\hj(z;\sigma,\tau)$,
the charge obeys the following differential equation
\begin{equation}
\frac{\d}{\d\tau}Q(z)\ =\
M(z;\sigma_0,\tau_0;\sigma+2\pi,\tau)
\left[\frac{\partial \hj}{\partial z}(z;\sigma,\tau),M(z;\sigma+2\pi,\tau;\sigma,\tau)\right]
M(z;\sigma,\tau;\sigma_0,\tau_0).
\end{equation}
In other words, the charge is conserved if the commutator on the right hand side vanishes.
The charges $Q(z)$ and the dual charges $\tilde Q(z)$ are related through \eqref{eq:LaxDuality},
though not in an obvious way.

Generically, the commutator can vanish only at specific values of $z$,
in particular at $z=\pm 1,\pm \im$, see \cite{Beisert:2005bm}.
The Lax connection at $z=\pm 1$ can be easily  integrated
\begin{equation}
 M(\pm 1;\sigma,\tau;\sigma_0,\tau_0)
  \ =\ P\exp\left(\int_{\sigma_0,\tau_0}^{\sigma,\tau} j\right)
  \ =\ g(\sigma,\tau)^{-1}g(\sigma_0,\tau_0).
\end{equation}
In particular, due to the assumed periodicity of $g$, one finds
\begin{equation}
 M(\pm 1;\sigma+2\pi,\tau;\sigma,\tau)\ =\ g(\sigma+2\pi,\tau)^{-1}g(\sigma,\tau)\ =\ 1,
\end{equation}
and therefore the charge at $z=\pm 1$ is manifestly conserved, $\frac{\d}{\d \tau} Q(\pm1)=0$.
This charge is the standard Noether charge for the $\mathfrak{so}(2,4)$ symmetry (see also
Eqs.~\eqref{eq:exp1}, \eqref{eq:exp2})
\begin{equation}
 Q(\pm 1)\ =\ \mp g(\sigma_0,\tau_0)^{-1} \left(\oint {*J_{_N}}\right) g(\sigma_0,\tau_0),
 \qquad\mbox{with}\qquad
 J_{_N}\ = \ g j_{(2)}g^{-1},
\end{equation}
i.e.\ $J_{_N}$ is the bosonic Noether current \eqref{eq:bosnicnoether}.

Consider now the dual charge $\tilde Q(z)$ at $z=\pm 1$. By similar arguments it is conserved if
$\tilde M(\pm 1;\sigma+2\pi,\tau;\sigma,\tau)$ commutes with $\partial_z \tilde \hj(\pm 1;\sigma,\tau)$.
However, this crucially depends on the periodicity of the dual coordinates $\tilde X^{\dot\alpha\beta}$
\begin{equation}
\begin{aligned}
\tilde M(\pm 1;\sigma+2\pi,\tau;\sigma,\tau)\ &=\ \tilde g(\sigma+2\pi,\tau)^{-1}\tilde g(\sigma,\tau)\\
 &=\ \exp\left[\im\tilde Y(\sigma)^{-1}\Big(\tilde X^{\dot\alpha\beta}(\sigma+2\pi)-\tilde
 X^{\dot\alpha\beta}(\sigma)\Big)P_{\beta\dot\alpha}\right]\!.
\end{aligned}
\end{equation}
In terms of the original coordinates this expression reads
\begin{equation}
 \tilde M(\pm 1;\sigma+2\pi,\tau;\sigma,\tau)\ =\
 \exp\left(-2\,Y\oint  {*\Omega(J_{_{N,K}}})\right)\!,
\end{equation}
where $J_{_{N,K}}$ is the projection of the Noether current \eqref{eq:bosnicnoether} along
the $K$-generator and $\Omega$ is the $\IZ_2$-automorphism defined in Eq.~\eqref{eq:z2auto}.
Thus the dual conformal symmetry acting on $\tilde X^{\dot\alpha\beta}$  coordinates
is manifest only if the Noether charge of the original model satisfies 
\begin{equation}\label{lasteq}
 \oint {*\Omega(J_{_{N,K}})}\ =\ 0, 
\end{equation}
i.e.\ the total momentum $\oint \d \sigma\, Y^2\partial_\tau  X_{\alpha \dot \beta}$  vanishes. 

To understand the meaning of this  conclusion, let us  recall  that
in the standard discussions of  T-duality  one usually assumes the compactness of  
the  isometry direction along which the duality is performed. Provided  the original $X$ 
and the dual $\tilde X$ coordinates  are periodic with radii $a$ and $\tilde a = \frac{\alpha'}{a}$,  
the T-duality is then a symmetry of the spectrum of  underlying conformal field
theory: it   interchanges  the Kaluza-Klein momenta with the winding mode numbers. 
Viewing  the  non-compact isometry case  as a limit  of the compact one means that to preserve
this  symmetry one  may assume that the dual coordinate is compactified on a circle of
vanishing radius,  $\tilde a \to 0$: then all  finite-mass momentum modes are mapped into 
finite-mass winding modes (see the  second  reference in \cite{Buscher:1987qj,Rocek:1991ps}).
A possible  alternative is to restrict consideration  to a subsector of 
states that do not carry momentum in the non-compact  isometric 
$X$ direction; then their duals  are not required to  have a winding in $\tilde X$ and thus 
$\tilde X$ may also be assumed to be non-compact. 
Indeed, Eq.~\eqref{lasteq} may be interpreted as such  zero $X$ momentum or zero $\tilde X$ winding 
condition.

Since the T-duality along all the four translational isometries of the AdS$_5$ space  acts also on the 
time direction,  it is not clear, even assuming the above zero-momentum condition, if 
this duality may have some useful implications for the  closed string 
spectrum of the superstring theory. Given a close  relation  between  the T-self-duality
of the AdS$_5\times S^5$  sigma model and  its integrability that we uncovered above, one may still 
expect some connection between the duality and the closed string spectrum, but that probably requires 
a certain complexification of the set of charges that label string states (in addition to a constraint
on their values implied by the above discussion).


\bibliographystyle{nb}
\bibliography{duallax}

\begin{thebibliography}{10}
\ifx\href\asklfhas\newcommand{\href}[2]{#2}\fi
\ifx\arxivref\asklfhas\newcommand{\arxivref}[2]{\href{http://arxiv.org/abs/#1}%
{#2}}\fi
\ifx\doiref\asklfhas\newcommand{\doiref}[2]{\href{http://dx.doi.org/#1}{#2}}\fi
\raggedright
\small
\parskip 0pt

\bibitem{Luscher:1977rq}
M.~L{\"u}scher and K.~Pohlmeyer,
\textit{``{Scattering of Massless Lumps and Nonlocal Charges in the
  Two-Dimensional Classical Nonlinear Sigma Model}''},
\textsf{\doiref{10.1016/0550-3213(78)90049-4}{Nucl.~Phys.~B137,~46~(1978)}}.
%
\bibitem{Mandal:2002fs}
G.~Mandal, N.~V.~Suryanarayana and S.~R.~Wadia,
\textit{``{Aspects of semiclassical strings in $AdS_5$}''},
\textsf{\doiref{10.1016/S0370-2693(02)02424-3}{Phys.~Lett.~B543,~81~(2002)}},
\texttt{\arxivref{hep-th/0206103}{hep-th/0206103}}.
%
\bibitem{Metsaev:1998it}
R.~R.~Metsaev and A.~A.~Tseytlin,
\textit{``{Type IIB superstring action in $AdS_5\times S^5$ background}''},
\textsf{\doiref{10.1016/S0550-3213(98)00570-7}{Nucl.~Phys.~B533,~109~(1998)}},
\texttt{\arxivref{hep-th/9805028}{hep-th/9805028}}.
%
\bibitem{Bena:2003wd}
I.~Bena, J.~Polchinski and R.~Roiban,
\textit{``{Hidden symmetries of the $AdS_5\times S^5$ superstring}''},
\textsf{\doiref{10.1103/PhysRevD.69.046002}{Phys.~Rev.~D69,~046002~(2004)}},
\texttt{\arxivref{hep-th/0305116}{hep-th/0305116}}.
%
\bibitem{Zakharov:1973pp}
V.~E.~Zakharov and A.~V.~Mikhailov,
\textit{``{Relativistically Invariant Two-Dimensional Models in Field Theory
  Integrable by the Inverse Problem Technique. (In Russian)}''},
\textsf{Sov.~Phys.~JETP~47,~1017~(1978)}.
%
\bibitem{Nappi:1979ig}
C.~R.~Nappi,
\textit{``{Some properties of an analog of the nonlinear sigma model}''},
\textsf{\doiref{10.1103/PhysRevD.21.418}{Phys.~Rev.~D21,~418~(1980)}}.
%
\bibitem{Fridling:1983ha}
B.~E.~Fridling and A.~Jevicki,
\textit{``{Dual representations and ultraviolet divergences in nonlinear sigma
  models}''},
\textsf{\doiref{10.1016/0370-2693(84)90987-0}{Phys.~Lett.~B134,~70~(1984)}}.
%
\bibitem{Fradkin:1984ai}
E.~S.~Fradkin and A.~A.~Tseytlin,
\textit{``{Quantum equivalence of dual field theories}''},
\textsf{\doiref{10.1016/0003-4916(85)90225-8}{Ann.~Phys.~162,~31~(1985)}}.
%
\bibitem{Buscher:1987qj}
T.~H.~Buscher,
\textit{``{Path Integral Derivation of Quantum Duality in Nonlinear Sigma
  Models}''},
\textsf{\doiref{10.1016/0370-2693(88)90602-8}{Phys.~Lett.~B201,~466~(1988)}}.
%
\bibitem{Rocek:1991ps}
M.~Ro\v{c}ek and E.~P.~Verlinde,
\textit{``{Duality, quotients, and currents}''},
\textsf{\doiref{10.1016/0550-3213(92)90269-H}{Nucl.~Phys.~B373,~630~(1992)}},
\texttt{\arxivref{hep-th/9110053}{hep-th/9110053}}.
%
\bibitem{Alvarez:1995zr}
E.~\'Alvarez, L.~\'Alvarez-Gaum\'e and I.~Bakas,
\textit{``{T duality and space-time supersymmetry}''},
\textsf{\doiref{10.1016/0550-3213(95)00566-8}{Nucl.~Phys.~B457,~3~(1995)}},
\texttt{\arxivref{hep-th/9507112}{hep-th/9507112}}.
%
\bibitem{Sfetsos:1995ac}
K.~Sfetsos,
\textit{``{Duality and Restoration of Manifest Supersymmetry}''},
\textsf{\doiref{10.1016/0550-3213(96)00005-3}{Nucl.~Phys.~B463,~33~(1996)}},
\texttt{\arxivref{hep-th/9510034}{hep-th/9510034}}.
%
\bibitem{Curtright:1996ig}
T.~Curtright, T.~Uematsu and C.~K.~Zachos,
\textit{``{Geometry and Duality in Supersymmetric sigma-Models}''},
\textsf{\doiref{10.1016/0550-3213(96)00138-1}{Nucl.~Phys.~B469,~488~(1996)}},
\texttt{\arxivref{hep-th/9601096}{hep-th/9601096}}.
%
\bibitem{Kallosh:1998ji}
R.~Kallosh and A.~A.~Tseytlin,
\textit{``{Simplifying superstring action on $AdS_5\times S^5$}''},
\textsf{\doiref{10.1088/1126-6708/1998/10/016}{JHEP~9810,~016~(1998)}},
\texttt{\arxivref{hep-th/9808088}{hep-th/9808088}}.
%
\bibitem{Alday:2007hr}
L.~F.~Alday and J.~M.~Maldacena,
\textit{``{Gluon scattering amplitudes at strong coupling}''},
\textsf{\doiref{10.1088/1126-6708/2007/06/064}{JHEP~0706,~064~(2007)}},
\texttt{\arxivref{0705.0303}{arxiv:0705.0303}}.
%
\bibitem{Alday:2007he}
L.~F.~Alday and J.~Maldacena,
\textit{``{Comments on gluon scattering amplitudes via AdS/CFT}''},
\textsf{\doiref{10.1088/1126-6708/2007/11/068}{JHEP~0711,~068~(2007)}},
\texttt{\arxivref{0710.1060}{arxiv:0710.1060}}.
%
\bibitem{Kruczenski:2007cy}
M.~Kruczenski, R.~Roiban, A.~Tirziu and A.~A.~Tseytlin,
\textit{``{Strong-coupling expansion of cusp anomaly and gluon amplitudes from
  quantum open strings in $AdS_5\times S^5$}''},
\textsf{\doiref{10.1016/j.nuclphysb.2007.09.005}{Nucl.~Phys.~B791,~93~(2008)}},
\texttt{\arxivref{0707.4254}{arxiv:0707.4254}}.
%
\bibitem{Drummond:2006rz}
J.~M.~Drummond, J.~Henn, V.~A.~Smirnov and E.~Sokatchev,
\textit{``{Magic identities for conformal four-point integrals}''},
\textsf{\doiref{10.1088/1126-6708/2007/01/064}{JHEP~0701,~064~(2007)}},
\texttt{\arxivref{hep-th/0607160}{hep-th/0607160}}.
%
\bibitem{Drummond:2007aua}
J.~M.~Drummond, G.~P.~Korchemsky and E.~Sokatchev,
\textit{``{Conformal properties of four-gluon planar amplitudes and Wilson
  loops}''},
\textsf{\doiref{10.1016/j.nuclphysb.2007.11.041}{Nucl.~Phys.~B795,~385~(2008)}%
},
\texttt{\arxivref{0707.0243}{arxiv:0707.0243}}.
%
\bibitem{Drummond:2007cf}
J.~M.~Drummond, J.~Henn, G.~P.~Korchemsky and E.~Sokatchev,
\textit{``{On planar gluon amplitudes/Wilson loops duality}''},
\textsf{\doiref{10.1016/j.nuclphysb.2007.11.007}{Nucl.~Phys.~B795,~52~(2008)}},
\texttt{\arxivref{0709.2368}{arxiv:0709.2368}}.
%
\bibitem{Drummond:2007au}
J.~M.~Drummond, J.~Henn, G.~P.~Korchemsky and E.~Sokatchev,
\textit{``{Conformal Ward identities for Wilson loops and a test of the duality
  with gluon amplitudes}''},
\texttt{\arxivref{0712.1223}{arxiv:0712.1223}}.
%
\bibitem{Drummond:2008aq}
J.~M.~Drummond, J.~Henn, G.~P.~Korchemsky and E.~Sokatchev,
\textit{``{Hexagon Wilson loop = six-gluon MHV amplitude}''},
\texttt{\arxivref{0803.1466}{arxiv:0803.1466}}.
%
\bibitem{Bern:2006ew}
Z.~Bern, M.~Czakon, L.~J.~Dixon, D.~A.~Kosower and V.~A.~Smirnov,
\textit{``{The Four-Loop Planar Amplitude and Cusp Anomalous Dimension in
  Maximally Supersymmetric Yang--Mills Theory}''},
\textsf{\doiref{10.1103/PhysRevD.75.085010}{Phys.~Rev.~D75,~085010~(2007)}},
\texttt{\arxivref{hep-th/0610248}{hep-th/0610248}}.
%
\bibitem{Bern:2007ct}
Z.~Bern, J.~J.~M.~Carrasco, H.~Johansson and D.~A.~Kosower,
\textit{``{Maximally supersymmetric planar Yang--Mills amplitudes at five
  loops}''},
\textsf{\doiref{10.1103/PhysRevD.76.125020}{Phys.~Rev.~D76,~125020~(2007)}},
\texttt{\arxivref{0705.1864}{arxiv:0705.1864}}.
%
\bibitem{Bern:2008ap}
Z.~Bern et~al.,
\textit{``{The Two-Loop Six-Gluon MHV Amplitude in Maximally Supersymmetric
  Yang-Mills Theory}''},
\textsf{\doiref{10.1103/PhysRevD.78.045007}{Phys.~Rev.~D78,~045007~(2008)}},
\texttt{\arxivref{0803.1465}{arxiv:0803.1465}}.
%
\bibitem{Komargodski:2008wa}
Z.~Komargodski,
\textit{``{On collinear factorization of Wilson loops and MHV amplitudes in
  {$\mathcal{N}=\mathord{}$4} SYM}''},
\textsf{\doiref{10.1088/1126-6708/2008/05/019}{JHEP~0805,~019~(2008)}},
\texttt{\arxivref{0801.3274}{arxiv:0801.3274}}.
%
\bibitem{Polyakov:1998ju}
A.~M.~Polyakov,
\textit{``{The wall of the cave}''},
\textsf{\doiref{10.1142/S0217751X99000324}{Int.~J.~Mod.~Phys.~A14,~645~(1999)}%
},
\texttt{\arxivref{hep-th/9809057}{hep-th/9809057}}.
%
\bibitem{Polyakov:2000fk}
A.~M.~Polyakov,
\textit{``{String theory as a universal language}''},
\textsf{\doiref{10.1134/1.1358479}{Phys.~Atom.~Nucl.~64,~540~(2001)}},
\texttt{\arxivref{hep-th/0006132}{hep-th/0006132}}.
%
\bibitem{Brandhuber:2007yx}
A.~Brandhuber, P.~Heslop and G.~Travaglini,
\textit{``{MHV Amplitudes in {$\mathcal{N}=\mathord{}$4} Super Yang--Mills and
  Wilson Loops}''},
\textsf{\doiref{10.1016/j.nuclphysb.2007.11.002}{Nucl.~Phys.~B794,~231~(2008)}%
},
\texttt{\arxivref{0707.1153}{arxiv:0707.1153}}.
%
\bibitem{McGreevy:2008zy}
J.~McGreevy and A.~Sever,
\textit{``{Planar scattering amplitudes from Wilson loops}''},
\textsf{\doiref{10.1088/1126-6708/2008/08/078}{JHEP~0808,~078~(2008)}},
\texttt{\arxivref{0806.0668}{arxiv:0806.0668}}.
%
\bibitem{Ricci:2007eq}
R.~Ricci, A.~A.~Tseytlin and M.~Wolf,
\textit{``{On T-Duality and Integrability for Strings on AdS Backgrounds}''},
\textsf{\doiref{10.1088/1126-6708/2007/12/082}{JHEP~0712,~082~(2007)}},
\texttt{\arxivref{0711.0707}{arxiv:0711.0707}}.
%
\bibitem{Hatsuda:2006ts}
M.~Hatsuda and S.~Mizoguchi,
\textit{``{Nonlocal charges of T-dual strings}''},
\textsf{\doiref{10.1088/1126-6708/2006/07/029}{JHEP~0607,~029~(2006)}},
\texttt{\arxivref{hep-th/0603097}{hep-th/0603097}}.
%
\bibitem{SokTalk}
E.~Sokatchev,
\textit{``Dual superconformal symmetry of scattering amplitudes in
  {$\mathcal{N}=\mathord{}$4} super-Yang--Mills''},
talk given at the conference ``Wonders of gauge theory and supergravity'',
  Paris, June 2008.
%
\bibitem{KorTalk}
G.~Korchemsky,
\textit{``Matching Wilson loops into scattering amplitudes in gauge
  theories''},
talk given at the conference ``Wonders of gauge theory and supergravity'',
  Paris, June 2008.
%
\bibitem{Drummond:2008vq}
J.~M.~Drummond, J.~Henn, G.~P.~Korchemsky and E.~Sokatchev,
\textit{``{Dual superconformal symmetry of scattering amplitudes in
  {$\mathcal{N}=\mathord{}$4} super-Yang--Mills theory}''},
\texttt{\arxivref{0807.1095}{arxiv:0807.1095}}.
%
\bibitem{BerTalk}
N.~Berkovits,
\textit{``Surprises in the $AdS_5\times S^5$ superstring''},
talk given at the conference ``Wonders of gauge theory and supergravity'',
  Paris, June 2008, based on work in preparation with J.~Maldacena.
%
\bibitem{Giveon:1994fu}
A.~Giveon, M.~Porrati and E.~Rabinovici,
\textit{``{Target space duality in string theory}''},
\textsf{\doiref{10.1016/0370-1573(94)90070-1}{Phys.~Rept.~244,~77~(1994)}},
\texttt{\arxivref{hep-th/9401139}{hep-th/9401139}}.
%
\bibitem{Alvarez:1994dn}
E.~\'Alvarez, L.~\'Alvarez-Gaum\'e and Y.~Lozano,
\textit{``{An introduction to T duality in string theory}''},
\textsf{\doiref{10.1016/0920-5632(95)00429-D}{Nucl.~Phys.~Proc.~Suppl.~41,~1~(%
1995)}},
\texttt{\arxivref{hep-th/9410237}{hep-th/9410237}}.
%
\bibitem{Kallosh:1998nx}
R.~Kallosh and J.~Rahmfeld,
\textit{``{The GS string action on $AdS_5\times S^5$}''},
\textsf{\doiref{10.1016/S0370-2693(98)01281-7}{Phys.~Lett.~B443,~143~(1998)}},
\texttt{\arxivref{hep-th/9808038}{hep-th/9808038}}.
%
\bibitem{Pesando:1998fv}
I.~Pesando,
\textit{``{A kappa gauge fixed type IIB superstring action on $AdS_5 \times
  S^5$}''},
\textsf{\doiref{10.1088/1126-6708/1998/11/002}{JHEP~9811,~002~(1998)}},
\texttt{\arxivref{hep-th/9808020}{hep-th/9808020}}.
%
\bibitem{Metsaev:2000yf}
R.~R.~Metsaev and A.~A.~Tseytlin,
\textit{``{Superstring action in $AdS_5\times S^5$: kappa-symmetry light cone
  gauge}''},
\textsf{\doiref{10.1103/PhysRevD.63.046002}{Phys.~Rev.~D63,~046002~(2001)}},
\texttt{\arxivref{hep-th/0007036}{hep-th/0007036}}.
%
\bibitem{Metsaev:2000yu}
R.~R.~Metsaev, C.~B.~Thorn and A.~A.~Tseytlin,
\textit{``{Light-cone superstring in AdS space-time}''},
\textsf{\doiref{10.1016/S0550-3213(00)00712-4}{Nucl.~Phys.~B596,~151~(2001)}},
\texttt{\arxivref{hep-th/0009171}{hep-th/0009171}}.
%
\bibitem{Roiban:2000yy}
R.~Roiban and W.~Siegel,
\textit{``{Superstrings on $AdS_5\times S^5$ supertwistor space}''},
\textsf{\doiref{10.1088/1126-6708/2000/11/024}{JHEP~0011,~024~(2000)}},
\texttt{\arxivref{hep-th/0010104}{hep-th/0010104}}.
%
\bibitem{Curtright:1994be}
T.~Curtright and C.~K.~Zachos,
\textit{``{Currents, charges, and canonical structure of pseudodual chiral
  models}''},
\textsf{\doiref{10.1103/PhysRevD.49.5408}{Phys.~Rev.~D49,~5408~(1994)}},
\texttt{\arxivref{hep-th/9401006}{hep-th/9401006}}.
%
\bibitem{Sarisaman:2007dm}
M.~Sarisaman,
\textit{``{Pseudoduality and Conserved Currents in Sigma Models}''},
\texttt{\arxivref{0704.3605}{arxiv:0704.3605}}.
%
\bibitem{Xiong:2007if}
C.-H.~Xiong,
\textit{``{The Hodge Dual Symmetry of the Green--Schwarz Superstring in $AdS_5
  \times S^5$}''},
\textsf{\doiref{10.1088/0253-6102/49/6/46}{Commun.~Theor.~Phys.~49,~1573~(2008%
)}},
\texttt{\arxivref{0705.3546}{arxiv:0705.3546}}.
%
\bibitem{Mikhailov:2005sy}
A.~Mikhailov,
\textit{``{A nonlocal Poisson bracket of the sine-Gordon model}''},
\texttt{\arxivref{hep-th/0511069}{hep-th/0511069}}.
%
\bibitem{Berkovits:1999zq}
N.~Berkovits, M.~Bershadsky, T.~Hauer, S.~Zhukov and B.~Zwiebach,
\textit{``{Superstring theory on $AdS_2\times S^2$ as a coset
  supermanifold}''},
\textsf{\doiref{10.1016/S0550-3213(99)00683-5}{Nucl.~Phys.~B567,~61~(2000)}},
\texttt{\arxivref{hep-th/9907200}{hep-th/9907200}}.
%
\bibitem{Polyakov:2004br}
A.~M.~Polyakov,
\textit{``{Conformal fixed points of unidentified gauge theories}''},
\textsf{\doiref{10.1142/S0217732304015129}{Mod.~Phys.~Lett.~A19,~1649~(2004)}},
\texttt{\arxivref{hep-th/0405106}{hep-th/0405106}}.
%
\bibitem{Alday:2005gi}
L.~F.~Alday, G.~Arutyunov and A.~A.~Tseytlin,
\textit{``{On integrability of classical superstrings in $AdS_5 \times
  S^5$}''},
\textsf{\doiref{10.1088/1126-6708/2005/07/002}{JHEP~0507,~002~(2005)}},
\texttt{\arxivref{hep-th/0502240}{hep-th/0502240}}.
%
\bibitem{Arutyunov:2004yx}
G.~Arutyunov and S.~Frolov,
\textit{``{Integrable Hamiltonian for classical strings on $AdS_5\times
  S^5$}''},
\textsf{\doiref{10.1088/1126-6708/2005/02/059}{JHEP~0502,~059~(2005)}},
\texttt{\arxivref{hep-th/0411089}{hep-th/0411089}}.
%
\bibitem{Grigoriev:2007bu}
M.~Grigoriev and A.~A.~Tseytlin,
\textit{``{Pohlmeyer reduction of $AdS_5\times S^5$ superstring sigma
  model}''},
\textsf{\doiref{10.1016/j.nuclphysb.2008.01.006}{Nucl.~Phys.~B800,~450~(2008)}%
},
\texttt{\arxivref{0711.0155}{arxiv:0711.0155}}.
%
\bibitem{Mikhailov:2007xr}
A.~Mikhailov and S.~Sch{\"a}fer-Nameki,
\textit{``{Sine-Gordon-like action for the Superstring in $AdS_5\times
  S^5$}''},
\textsf{\doiref{10.1088/1126-6708/2008/05/075}{JHEP~0805,~075~(2008)}},
\texttt{\arxivref{0711.0195}{arxiv:0711.0195}}.
%
\bibitem{Young:2005jv}
C.~A.~S.~Young,
\textit{``{Non-local charges, $Z_m$ gradings and coset space actions}''},
\textsf{\doiref{10.1016/j.physletb.2005.10.090}{Phys.~Lett.~B632,~559~(2006)}},
\texttt{\arxivref{hep-th/0503008}{hep-th/0503008}}.
%
\bibitem{Beisert:2005bm}
N.~Beisert, V.~A.~Kazakov, K.~Sakai and K.~Zarembo,
\textit{``{The algebraic curve of classical superstrings on $AdS_5\times
  S^5$}''},
\textsf{\doiref{10.1007/s00220-006-1529-4}{Commun.~Math.~Phys.~263,~659~(2006)%
}},
\texttt{\arxivref{hep-th/0502226}{hep-th/0502226}}.
%
\bibitem{Dolan:1983bp}
L.~Dolan,
\textit{``{Kac--Moody algebras and exact solvability in hadronic physics}''},
\textsf{\doiref{10.1016/0370-1573(84)90134-0}{Phys.~Rept.~109,~1~(1984)}}.
%
\bibitem{Kallosh:1998zx}
R.~Kallosh, J.~Rahmfeld and A.~Rajaraman,
\textit{``{Near horizon superspace}''},
\textsf{\doiref{10.1088/1126-6708/1998/09/002}{JHEP~9809,~002~(1998)}},
\texttt{\arxivref{hep-th/9805217}{hep-th/9805217}}.
%
\bibitem{Claus:1998yw}
P.~Claus and R.~Kallosh,
\textit{``{Superisometries of the $AdS \times S$ superspace}''},
\textsf{\doiref{10.1088/1126-6708/1999/03/014}{JHEP~9903,~014~(1999)}},
\texttt{\arxivref{hep-th/9812087}{hep-th/9812087}}.
%
\bibitem{Cvetic:1999zs}
M.~Cveti\v{c}, H.~Lu, C.~N.~Pope and K.~S.~Stelle,
\textit{``{T-duality in the Green--Schwarz formalism, and the massless/massive
  IIA duality map}''},
\textsf{\doiref{10.1016/S0550-3213(99)00740-3}{Nucl.~Phys.~B573,~149~(2000)}},
\texttt{\arxivref{hep-th/9907202}{hep-th/9907202}}.
%
\bibitem{Kulik:2000nr}
B.~Kulik and R.~Roiban,
\textit{``{T-duality of the Green--Schwarz superstring}''},
\textsf{\doiref{10.1088/1126-6708/2002/09/007}{JHEP~0209,~007~(2002)}},
\texttt{\arxivref{hep-th/0012010}{hep-th/0012010}}.
%
\bibitem{Hassan:2000kr}
S.~F.~Hassan,
\textit{``{Supersymmetry and the systematics of T-duality rotations in type-II
  superstring theories}''},
\textsf{\doiref{10.1016/S0920-5632(01)01539-0}{Nucl.~Phys.~Proc.~Suppl.~102,~7%
7~(2001)}},
\texttt{\arxivref{hep-th/0103149}{hep-th/0103149}}.
%
\bibitem{Alday:2005ww}
L.~F.~Alday, G.~Arutyunov and S.~Frolov,
\textit{``{Green--Schwarz strings in TsT-transformed backgrounds}''},
\textsf{\doiref{10.1088/1126-6708/2006/06/018}{JHEP~0606,~018~(2006)}},
\texttt{\arxivref{hep-th/0512253}{hep-th/0512253}}.
%
\bibitem{Schwarz:1992te}
A.~S.~Schwarz and A.~A.~Tseytlin,
\textit{``{Dilaton shift under duality and torsion of elliptic complex}''},
\textsf{\doiref{10.1016/0550-3213(93)90514-P}{Nucl.~Phys.~B399,~691~(1993)}},
\texttt{\arxivref{hep-th/9210015}{hep-th/9210015}}.
%
\bibitem{Arutyunov:2005nk}
G.~Arutyunov and M.~Zamaklar,
\textit{``{Linking B\"acklund and monodromy charges for strings on $AdS_5\times
  S^5$}''},
\textsf{\doiref{10.1088/1126-6708/2005/07/026}{JHEP~0507,~026~(2005)}},
\texttt{\arxivref{hep-th/0504144}{hep-th/0504144}}.
%
\bibitem{Alday:2008yw}
L.~F.~Alday and R.~Roiban,
\textit{``{Scattering Amplitudes, Wilson Loops and the String/Gauge Theory
  Correspondence}''},
\textsf{\doiref{10.1016/j.physrep.2008.08.002}{Phys.~Rept.~468,~153~(2008)}},
\texttt{\arxivref{0807.1889}{arxiv:0807.1889}}.
%
\bibitem{Berkovits:2008ic}
N.~Berkovits and J.~Maldacena,
\textit{``{Fermionic T-Duality, Dual Superconformal Symmetry, and the
  Amplitude/Wilson Loop Connection}''},
\textsf{\doiref{10.1088/1126-6708/2008/09/062}{JHEP~0809,~062~(2008)}},
\texttt{\arxivref{0807.3196}{arxiv:0807.3196}}.
%
\bibitem{Wess:1992cp}
J.~Wess and J.~Bagger,
\textit{``{Supersymmetry and supergravity}''},
Princeton University Press (1992),
Princeton, USA,
259p.
%
\end{thebibliography}
\end{document}